\def\IsExtended{1}
\if 1\IsExtended
\documentclass[envcountsame, a4paper]{article}
\else
\documentclass[orivec,envcountsame, a4paper]{llncs}
\fi

\newcommand{\ourtitle}{A Datalog Hammer for Supervisor Verification Conditions Modulo Simple Linear Arithmetic}

\usepackage{amsmath}

\usepackage{amsfonts}
\usepackage{amssymb}
\usepackage{amsthm}

\usepackage[utf8]{inputenc}
\usepackage{newtxtext,newtxmath} 
\DeclareMathAlphabet{\mathcal}{OMS}{cmsy}{m}{n} 
\usepackage{paralist} 
\usepackage{graphicx}

 \usepackage{comment}

\usepackage{url}
\usepackage[bookmarks=false,pdfpagelabels,bookmarksnumbered,bookmarksopen,bookmarksopenlevel=1,colorlinks=true,
citecolor=black, filecolor=black, linkcolor=black, menucolor=black,
urlcolor=black, pdftitle={\ourtitle}, pdfauthor={Martin Bromberger, Irina Dragoste, Rasha Faqeh, Christof Fetzer, Markus Kroetzsch, Christoph Weidenbach},
pdfkeywords={Datalog, arithmetic constraints, Bernays-Schoenfinkel first-order logic, SUPERLOG, SMT}]{hyperref}
\usepackage{listings} 
\lstdefinestyle{embedded}{
	numbers=none,
	frame=none,
	xleftmargin=0cm,
	backgroundcolor=\color{Lavender},
	framesep=1pt,
	aboveskip=3pt,
	belowskip=3pt,
    captionpos=b,                    
}
\lstdefinestyle{small}{
	basicstyle=\linespread{0.9}\footnotesize,
}

\mathchardef\hyphenmathcode=\mathcode`\-

\let\origlstlisting=\lstlisting
\let\endoriglstlisting=\endlstlisting
\renewenvironment{lstlisting}
    {\mathcode`\-=\hyphenmathcode
     \everymath{}\mathsurround=0pt\origlstlisting}
    {\endoriglstlisting}

\usepackage{xcolor} 

\usepackage{wrapfig} 

\usepackage[english]{babel}

\theoremstyle{definition}

\if 1\IsExtended

\newenvironment{prooftoggle}{\begin{proof}}{\end{proof}}
\newenvironment{extendedonly}{}{}
\excludecomment{notinextended}

\theoremstyle{definition}
\newtheorem{theorem}{Theorem}
\newtheorem{lemma}[theorem]{Lemma}

\newtheorem{corollary}[theorem]{Corollary}
\newtheorem{example}[theorem]{Example}
\newtheorem{definition}[theorem]{Definition}

\else

\excludecomment{prooftoggle}
\excludecomment{extendedonly}
\newenvironment{notinextended}{}{}

\usepackage[subtle]{savetrees}

\fi

\newcommand{\Real}{\mathbb{R}}

\newcommand{\Int}{\mathbb{Z}}

\newcommand{\preds}{\Pi}
\newcommand{\varset}{\mathcal{X}}

\newcommand{\inta}{\mathcal{A}}
\newcommand{\sigval}{\mathcal{A}}
\newcommand{\vars}{\ensuremath{\operatorname{vars}}}
\newcommand{\atoms}{\ensuremath{\operatorname{atoms}}}

\newcommand{\dom}{\operatorname{dom}}
\newcommand{\cdom}{\operatorname{codom}}
\newcommand{\comp}{\operatorname{comp}} 
\newcommand{\mGnd}{\operatorname{gnd}}
\newcommand{\mMGU}{\operatorname{mgu}} 

\newcommand{\ints}{\operatorname{ints}}

\newcommand{\dblquotes}{}

\newcommand{\intvp}{\mathcal{I}}

\newcommand{\idef}{\operatorname{idef}}
\newcommand{\iubnd}{\operatorname{iubd}}
\newcommand{\ilbnd}{\operatorname{ilbd}}
\newcommand{\tren}{\operatorname{tren}}
\newcommand{\tfacts}{\operatorname{tfacts}}
\newcommand{\pflat}{\operatorname{pflat}}
\newcommand{\rflat}{\operatorname{rflat}}

\newcommand{\elim}{\ensuremath{\operatorname{elim}}}

\newcommand{\LA}{\ensuremath{\operatorname{LA}}}
\newcommand{\LRA}{\ensuremath{\operatorname{LRA}}}
\newcommand{\SB}{\ensuremath{\operatorname{SLR}}}
\newcommand{\BS}{\ensuremath{\operatorname{BS}}}
\newcommand{\PP}{\ensuremath{\operatorname{PP}}}
\newcommand{\SP}{\ensuremath{\operatorname{P}}}
\newcommand{\LAOP}{\ensuremath{\operatorname{\triangleleft}}}

\newcommand{\HBS}{\ensuremath{\operatorname{HBS}}}

\newcommand{\Rulewerk}{Rulewerk}

\newcommand{\myparagraph}[1]{\smallskip \noindent{\textbf{#1}}}
\newcommand{\mysubparagraph}[1]{\smallskip{\emph{#1}}}

\newcommand{\figref}[1]{Fig.~\ref{fig:#1}}
\newcommand{\lstref}[1]{List.~\ref{lst:#1}}

\newcommand{\linref}[1]{line~\ref{line:#1}} 
\newcommand{\linreftwo}[2]{lines~\ref{line:#1}-\ref{line:#2}} 
	
\newcommand{\tspace}[1]{ 
\newcount\foo
\foo=#1
\loop
{\,}
\advance \foo -1
\ifnum \foo>0
\repeat
}

\definecolor{commentcolor}{gray}{0.5} 
\newcommand{\speccomment}[1]{\textcolor{commentcolor}{\emph{#1}}} 

\newcommand{\ie}{\emph{i.e.}, } 
\newcommand{\code}[1]{\texttt{#1}}

\title{\ourtitle}
\if 1\IsExtended
\author{Martin Bromberger\\ Max Planck Institute for Informatics\\ Saarland Informatics Campus,  Saarbr\"ucken, Germany\\  \and
  Irina Dragoste\\ TU Dresden, Dresden, Germany\\ \and
  Rasha Faqeh\\ TU Dresden, Dresden, Germany\\ \and
  Christof Fetzer\\ TU Dresden, Dresden, Germany\\ \and
  Markus Kr\"otzsch\\ TU Dresden, Dresden, Germany\\ \and
  Christoph Weidenbach \\ Max Planck Institute for Informatics\\ Saarland Informatics Campus, Saarbr\"ucken, Germany\\}
\else
\author{Martin~Bromberger\inst{1}\and Irina~Dragoste\inst{2} \and Rasha~Faqeh\inst{2} \and Christof~Fetzer\inst{2} \and Markus~Kr\"otzsch\inst{2} \and Christoph~Weidenbach\inst{1}}

\institute{Max Planck Institute for Informatics, Saarland Informatics Campus, Saarbr\"ucken\and TU Dresden, Dresden, Germany}
\fi

\begin{document}

\maketitle

\begin{abstract}
  The Bernays-Schönfinkel first-order logic fragment over simple linear real arithmetic
  constraints BS(SLR) is known to be decidable. We prove that BS(SLR) clause sets with both
  universally and existentially quantified verification conditions (conjectures) can be translated into BS(SLR) clause
  sets over a finite set of first-order constants. For the Horn case, we provide a 
  Datalog hammer preserving validity and satisfiability. A toolchain from the BS(LRA) prover SPASS-SPL
  to the Datalog reasoner VLog establishes an effective way of deciding verification
  conditions in the Horn fragment. This is exemplified by the verification of supervisor code for a lane change
  assistant in a car and of an electronic control unit for a supercharged combustion engine.
\end{abstract}

\section{Introduction} \label{sec:intro}

Modern dynamic dependable systems (e.g., autonomous driving) continuously update software components to fix bugs and to introduce new features.
However, the safety requirement of such systems demands software to be safety certified before it can be used, which is typically a lengthy process
that hinders the dynamic update of software.
We adapt the \emph{continuous certification} approach \cite{FaqehFH0KKSW20} of variants of safety critical software components
using a \emph{supervisor} that guarantees important aspects through challenging, see \figref{supervisor-arch}.
Specifically, multiple processing units run in parallel -- \emph{certified} and \emph{updated not-certified} variants that produce output as \emph{suggestions} and \emph{explications}.
The supervisor compares the behavior of variants and analyses their explications. The supervisor itself consists of a rather small set
of rules that can be automatically verified and run by a \emph{reasoner}.
The reasoner helps the supervisor to check if the output of an updated variant is in agreement with the output of a respective certified variant. 
The absence of discrepancy between the two variants for a long-enough period of running both variants in parallel
allows to dynamically certify it as a safe software variant.

\begin{extendedonly}
\begin{figure}[t]
	\begin{center}
		\includegraphics[scale=0.6]{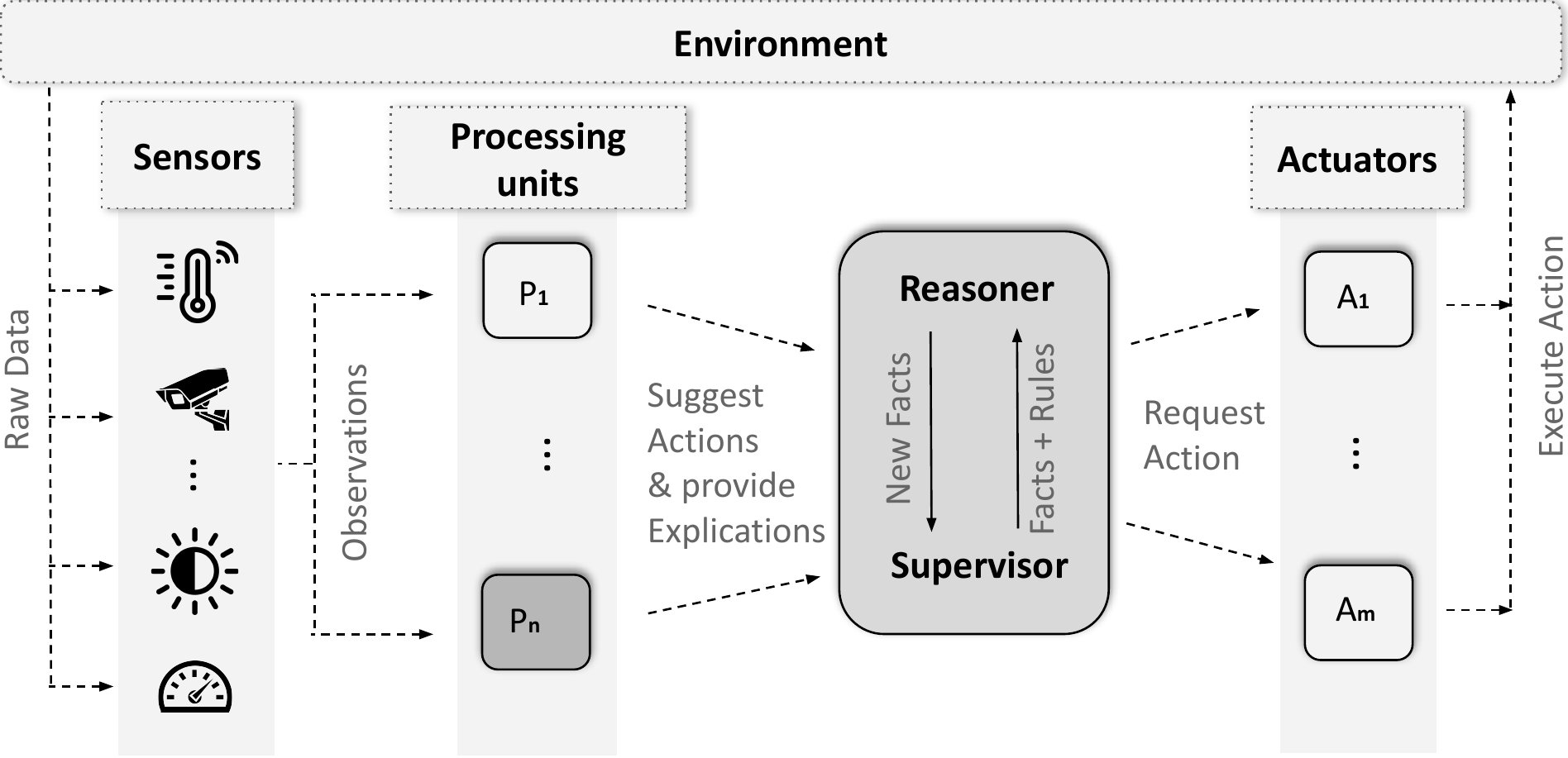}
        \caption{The supervisor architecture.}
        \label{fig:supervisor-arch}
    \end{center}
\end{figure}
\end{extendedonly}
\begin{notinextended}
\begin{figure}[h]
	\begin{center}
	     \vspace{-20pt}
		\includegraphics[scale=0.5]{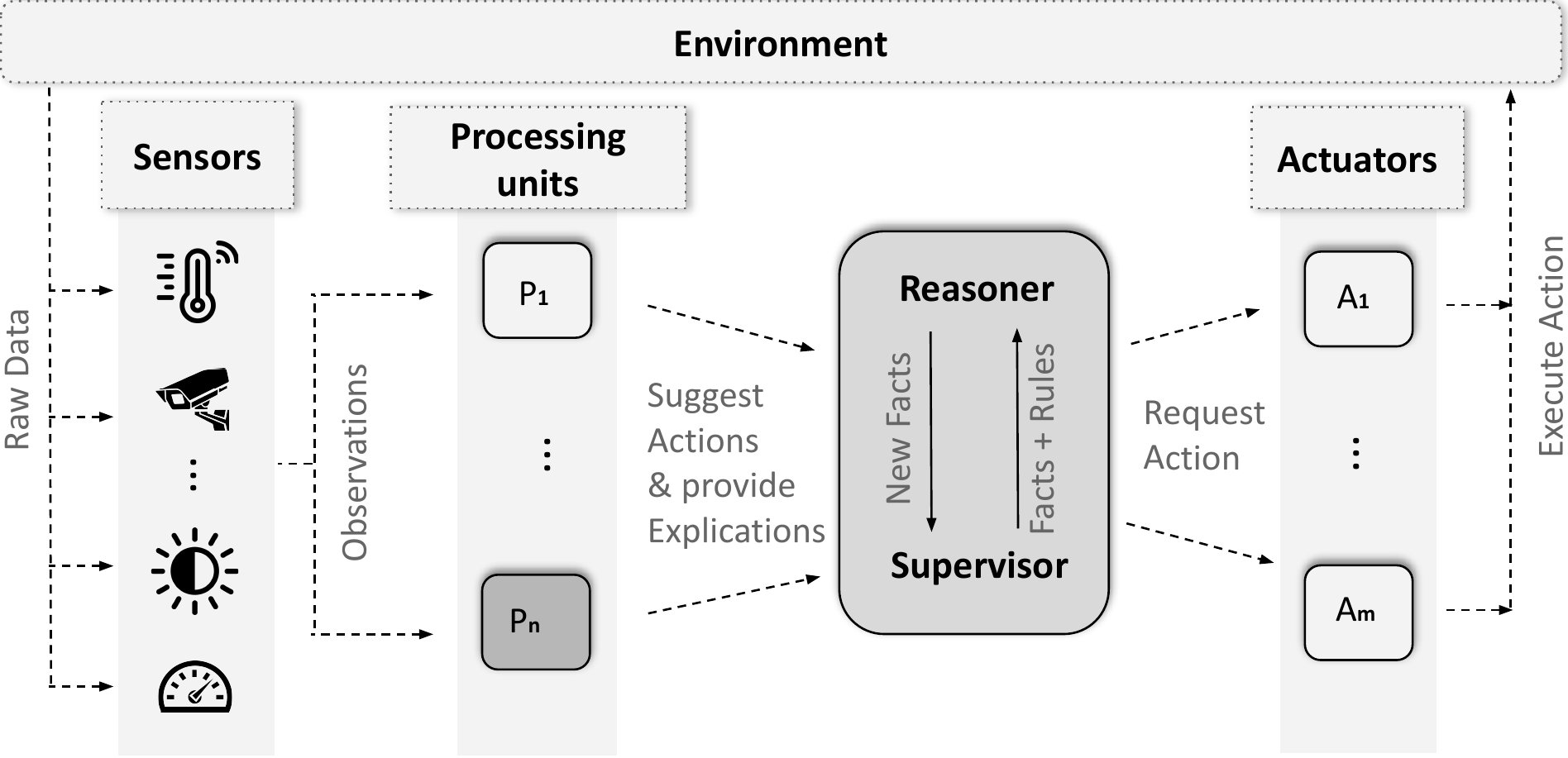}
        \caption{The supervisor architecture.}
        \label{fig:supervisor-arch}
    \end{center}
\end{figure}
\end{notinextended}

While supervisor safety
conditions formalized as existentially quantified properties can often
already be automatically verified,
conjectures about invariants formalized as universally quantified properties are a further challenge. In this
paper we show that supervisor safety conditions and invariants can be automatically proven
by a Datalog hammer. Analogous to the Sledgehammer project~\cite{BoehmeNipkow10} of Isabelle~\cite{NipkowEtAl02} translating higher-order logic
conjectures to first-order logic (modulo theories) conjectures, our Datalog hammer translates first-order
Horn logic modulo arithmetic conjectures into pure Datalog programs, equivalent to Horn Bernays-Sch\"onfinkel clause fragment, called $\HBS$.

More concretely, the underlying logic for both formalizing supervisor behavior and formulating
conjectures is the hierarchic combination
of the Bernays-Sch\"onfinkel first-order fragment with real linear arithmetic, $\BS(\LRA)$,
also called \emph{Superlog} for Supervisor Effective Reasoning Logics~\cite{FaqehFH0KKSW20}.
Satisfiability of $\BS(\LRA)$ clause sets is undecidable~\cite{Downey1972,HorbachEtAl17ARXIV}, in general, however, the restriction
to simple linear real arithmetic $\BS(\SB)$ yields a decidable fragment~\cite{GeMoura09,HorbachEtAl17CADE}.
Our first contribution is decidability of $\BS(\SB)$ with respect to universally quantified conjectures, Section~\ref{sec:theory}, Lemma~\ref{lem:verificationequiv}.

Inspired by the test point method for quantifier elimination in arithmetic~\cite{LoosWeispfenning93} we
show that instantiation with a finite number of first-order constants is sufficient to decide
whether
a universal/existential conjecture is a consequence of a $\BS(\SB)$ clause set.

For our experiments of the test point approach we consider two case studies: verification
conditions for a supervisor taking care of multiple software variants of a lane change assistant in a car and a supervisor 
for a supercharged combustion engine, also called an ECU for Electronical Control Unit. The supervisors in both cases
are formulated by $\BS(\SB)$ Horn clauses, the $\HBS(\SB)$ fragment.
Via our test point technique they are
translated together with the verification conditions to Datalog~\cite{Alice} ($\HBS$).
The translation is implemented in our Superlog reasoner SPASS-SPL. The resulting Datalog
clause set is eventually explored by the Datalog engine VLog~\cite{Rulewerk2019}.
This
hammer constitutes a decision procedure for both universal and existential conjectures.
The results of our experiments show that we can verify non-trivial existential
and universal conjectures in the range of seconds while state-of-the-art solvers
cannot solve all problems in reasonable time. This constitutes our second contribution, Section~\ref{sec:experiments}.

\myparagraph{Related Work:} Reasoning about $\BS(\LRA)$ clause sets is supported by SMT (Satisfiability Modulo Theories)~\cite{NieuwenhuisEtAl06,MouraBjorner11}.
In general, SMT comprises the combination of a number of theories beyond $\LRA$ such as arrays, lists, strings, or bit vectors.
While SMT is a decision procedure for the $\BS(\LRA)$ ground case, universally quantified variables can be considered by instantiation~\cite{ReynoldsEtAl18}.
Reasoning by instantiation does result in a refutationally complete procedure for $\BS(\SB)$, but not in a decision procedure. The Horn
fragment $\HBS(\LRA)$ out of $\BS(\LRA)$ is receiving additional attention~\cite{GrebenshchikovEtAl12,BjornerEtAl15}, because it is well-suited for 
software analysis and verification. Research in this direction also goes beyond the theory of $\LRA$ and considers minimal model semantics in addition, but is restricted
to existential conjectures.
Other research focuses on universal conjectures, but over non-arithmetic theories, e.g., invariant checking for array-based systems~\cite{CimattiGR21}
or considers abstract dedidability criteria incomparable with the $\HBS(\LRA)$ class~\cite{Ranise12}.
Hierarchic superposition~\cite{BachmairGanzingerEtAl94} and Simple Clause Learning over Theories~\cite{BrombergerFW21} (SCL(T)) are both
refutationally complete for $\BS(\LRA)$. While SCL(T) can be immediately turned into a decision procedure for even larger fragments than $\BS(\SB)$~\cite{BrombergerFW21}, hierarchic
superposition needs to be refined by specific strategies or rules to become a decision procedure already because of the Bernays-Sch\"onfinkel part~\cite{HillenbrandWeidenbach13}.
Our Datalog hammer translates $\HBS(\SB)$ clause sets with both existential and universal conjectures into $\HBS$ clause sets which
are also subject to first-order theorem proving. Instance generating approaches such as  iProver~\cite{Korovin08} are a decision procedure
for this fragment, whereas superposition-based~\cite{BachmairGanzingerEtAl94} first-order provers
such as E~\cite{SchulzEtAl19}, SPASS~\cite{WeidenbachEtAlSpass2009}, Vampire~\cite{RiazanovVoronkov02},
have additional mechanisms implemented to decide $\HBS$.
In our experiments, Section~\ref{sec:experiments}, we will discuss the differences between all these approaches on a number of benchmark examples in more detail.

The paper is organized as follows: after a section on preliminaries, Section~\ref{sec:prelims}, we present the theory
of our new Datalog hammer in Section~\ref{sec:theory}. Section~\ref{sec:super} introduces our two case studies followed by
experiments on respective verification conditions, Section~\ref{sec:experiments}. The paper ends with a discussion of
the obtained results and directions for future work, Section~\ref{sec:conclusion}.
\begin{notinextended}
    Binaries of our tools, all benchmark problems, and an extended version of this paper including all proofs
    can be found under \url{https://github.com/knowsys/eval-datalog-arithmetic}.
\end{notinextended}
\begin{extendedonly}
  This paper is an extended version of~\cite{BrombergerEtAl21FROCOS}. Binaries of our tools, and all benchmark problems, 
  can be found under \url{https://github.com/knowsys/eval-datalog-arithmetic}.
\end{extendedonly}


\section{Preliminaries}\label{sec:prelims}

We briefly recall the basic logical formalisms and notations we build upon.
We use a standard first-order language with \emph{constants} (denoted $a, b, c$), without non-constant function symbols,
\emph{variables} (denoted $w,x,y,z$), and \emph{predicates} (denoted $P, Q, R$)
of some fixed \emph{arity}.
\emph{Terms} (denoted $t,s$) are variables or constants.
We write $\bar{x}$ for a vector of variables, $\bar{a}$ for a vector of constants, and so on.
An \emph{atom} (denoted $A, B$) is an expression $P(\bar{t})$ for a predicate $P$ of arity $n$ and a term list $\bar{t}$
of length $n$.
A \emph{positive literal} is an atom $A$ and a \emph{negative literal} is a negated atom $\neg A$.
We define $\comp(A)=\neg A$, $\comp(\neg A)=A$, $|A|=A$ and $|\neg A|=A$.
Literals are usually denoted $L, K, H$.

A \emph{clause} is a disjunction of literals, where all variables are assumed to be universally
quantified. $C, D$ denote clauses, and $N$ denotes a clause set.
We write $\atoms(X)$ for the set of atoms in a clause or clause set $X$.
A clause is \emph{Horn} if it contains at most one positive literal, and
a \emph{unit clause} if it has exactly one literal.
A clause $A_1\vee \ldots\vee A_n\vee \neg B_1\vee\ldots\vee \neg B_m$ can be written as an implication
$A_1\wedge \ldots\wedge A_n\to B_1\vee\ldots\vee B_m$, still omitting universal quantifiers.
If $Y$ is a term, formula, or a set thereof, $\vars(Y)$ denotes the set of all variables in $Y$,
and $Y$ is \emph{ground} if $\vars(Y)=\emptyset$.
A \emph{fact} is a ground unit clause with a positive literal.

\myparagraph{Datalog and the Bernays-Sch\"onfinkel Fragment:}
The \emph{Bernays-Schönfinkel fragment} (\BS) comprises all sets of clauses.
The more general form of \BS{} in first-order logic allows arbitrary \emph{formulas} over atoms,
i.e., arbitrary Boolean connectives and leading existential quantifiers. 
However, both can be polynomially removed with common 
syntactic transformations while preserving satisfiability and all entailments that do not refer to auxiliary
constants and predicates introduced in the transformation~\cite{NonnengartW01}.
Sometimes, we still refer explicitly to formulas when it is more beneficial to apply these transformations after some other processing steps.
\BS{} theories in our sense are also known as \emph{disjunctive Datalog programs} \cite{DisjunctiveDatalog97},
specifically when written as implications.
A set of Horn clauses is also called a \emph{Datalog program}.
(Datalog is sometimes viewed as
a second-order language. We are only interested in query answering, which can equivalently be viewed as first-order entailment
or second-order model checking \cite{Alice}.) 
Again, it is common to write clauses as implications in this case.

Two types of \emph{conjectures}, i.e., formulas we want
to prove as consequences of a clause set, are of particular interest:
\emph{universal} conjectures $\forall \bar{x} \phi$ and \emph{existential} conjectures $\exists \bar{x} \phi$,
where $\phi$ is any Boolean combination of $\BS$ atoms that only uses variables in $\bar{x}$.

A \emph{substitution} $\sigma$ is a function from variables to terms with a finite domain 
$\dom(\sigma) = \{ x \mid x\sigma \neq x\}$ and codomain $\cdom(\sigma) = \{x\sigma\mid x\in\dom(\sigma)\}$.
We denote substitutions by $\sigma, \delta, \rho$.
The application of substitutions is often written postfix, as in $x\sigma$, and is homomorphically extended to
terms, atoms, literals, clauses, and quantifier-free formulas.
A substitution $\sigma$ is \emph{ground} if $\cdom(\sigma)$ is ground.
Let $Y$ denote some term, literal, clause, or clause set.
$\sigma$ is a \emph{grounding} for $Y$ if $Y\sigma$ is ground, and $Y\sigma$ is a
\emph{ground instance} of $Y$ in this case.
We denote by $\mGnd(Y)$ the set of all ground instances of $Y$, 
and by $\mGnd_B(Y)$ the set of all ground instances over a given set of constants $B$.
The \emph{most general unifier} $\mMGU(Z_1,Z_2)$ of two terms/atoms/literals $Z_1$ and $Z_2$
is defined as usual, and we assume that it does not introduce fresh variables and is idempotent.

We assume a standard first-order logic model theory, and write
$\sigval \models \phi$ if an interpretation $\sigval$ satisfies a first-order formula $\phi$.
A formula $\psi$ is a logical consequence of $\phi$, written $\phi\models\psi$, if
$\sigval\models\psi$ for all $\sigval$ such that $\sigval\models\phi$.
Sets of clauses are semantically treated as conjunctions of clauses with all variables quantified
universally.

\myparagraph{$\BS$ with Linear Arithmetic:}
The extension of $\BS$ with linear arithmetic over real numbers, $\BS(\LRA)$,
is the basis for the formalisms studied in this paper.
For simplicity, we assume a one-sorted extension where all terms in $\BS(\LRA)$ are of arithmetic sort $\LA$, i.e., represent numbers.
The language includes free first-order logic constants that are eventually interpreted by real numbers, but we only
consider initial clause sets without such constants, called \emph{pure} clause sets.
Satisfiability of pure $\BS(\LRA)$ clause sets is semi-decidable, e.g., using \emph{hierarchic superposition} \cite{BachmairGanzingerEtAl94}
or \emph{SCL(T)}~\cite{BrombergerFW21}.
Impure $\BS(\LRA)$ is no longer compact and satisfiability becomes undecidable, 
but it can be made decidable when restricting to ground clause sets~\cite{BrombergerEtAl2020arxiv}, which
is the result of our grounding hammer.

\begin{example}\label{ex_bs_lra_ecu}
The following $\BS(\LRA)$ clause from our ECU case study compares the
values of speed ($\text{Rpm}$) and pressure ($\text{KPa}$) with entries in an ignition
table ($\text{IgnTable}$) to derive the basis of the current ignition value ($\text{IgnDeg1}$):
\begin{align}\begin{split}
  &x_1 < 0 \;\lor\; x_1\geq 13 \;\lor\; x_2 < 880  \;\lor\; x_2\geq 1100 \;\lor\; \neg \text{KPa}(x_3, x_1) \;\lor{} \\
  &  \neg \text{Rpm}(x_4, x_2) \;\lor\; \neg \text{IgnTable}(0,13,880,1100,z)  \;\lor\; \text{IgnDeg1}(x_3, x_4, x_1, x_2, z)
\end{split}\label{eq_bs_lra_ecu}
\end{align}
\end{example}

Terms of sort $\LA$ are constructed from a set $\varset$ of \emph{variables}, a set of \emph{first-order arithmetic constants},
the set of integer constants $c\in\Int$, and binary function symbols $+$ and $-$ (written infix).
Atoms in $\BS(\LRA)$ are either \emph{first-order atoms} (e.g., $\text{IgnTable}(0,13,880,1100,z)$) or
\emph{(linear) arithmetic atoms} (e.g., $x_2 < 880$).
Arithmetic atoms may use the predicates $\leq, <, \neq, =, >, \geq$, which are written infix and have the expected
fixed interpretation. Predicates used in first-order atoms are called \emph{free}.
\emph{First-order literals} and related notation is defined as before.
\emph{Arithmetic literals} coincide with arithmetic atoms, since
the arithmetic predicates are closed under negation, e.g., $\comp(x_2\geq 1100)=x_2<1100$.

$\BS(\LRA)$ clauses and conjectures are defined as for \BS{} but using $\BS(\LRA)$ atoms.
We often write clauses in the form $\Lambda \parallel C$ where $C$ is a clause solely built of free first-order literals
and $\Lambda$ is a multiset of $\LRA$ atoms. The semantics of $\parallel$ is implication where $\Lambda$ denotes a conjunction, e.g.,
the clause $x >1 \lor y \neq 5 \lor \neg Q(x) \lor R(x, y)$ is also written $x\leq 1, y = 5 || \neg Q(x) \lor R(x,y)$.
For $Y$ a term, literal, or clause, we write $\ints(Y)$ for the set of all integers that occur in $Y$. 

A clause or clause set is \emph{pure} if it does not contain first-order arithmetic constants,
and it is \emph{abstracted} if its first-order literals contain only variables.
Every clause $C$ is equivalent to an abstracted clause that is obtained by replacing each non-variable term $t$
that occurs in a first-order atom by a fresh variable $x$ while adding an arithmetic atom $x\neq t$ to $C$.
We asssume abstracted clauses for theory development, but we prefer non-abstracted
clauses in examples for readability,e.g., a fact $P(3,5)$ is considered in the development
of the theory as the clause $x=3, x=5 || P(x,y)$, this is important when collecting the
necessary test points.

The semantics of $\BS(\LRA)$ is based on the standard model $\sigval^{\LRA}$ of linear arithmetic,
which has the domain $\LA^{\sigval^{\LRA}} = \Real$ and which interprets all arithmetic predicates and functions in the
usual way.
An interpretation of $\BS(\LRA)$ coincides with $\sigval^{\LRA}$ on arithmetic predicates and functions, and freely interprets
free predicates and first-order arithmetic constants. For pure clause sets this is well-defined~\cite{BachmairGanzingerEtAl94}.
Logical satisfaction and entailment is defined as usual, and uses similar notation as for \BS.

\myparagraph{Simpler Forms of Linear Arithmetic:} \label{sec:bsslr}
The main logic studied in this paper is obtained by restricting $\BS(\LRA)$ to 
a simpler form of linear arithmetic. We first introduce a simpler logic $\BS(\SB)$
as a well-known fragment of $\BS(\LRA)$ for which satisfiability is decidable~\cite{GeMoura09,HorbachEtAl17CADE},
and then present the generalization $\BS(\LRA)\PP$ of this formalism that we will use.

\begin{definition}
The \emph{Bernays-Schönfinkel fragment over simple linear arithmetic}, $\BS(\SB)$, is a subset of
$\BS(\LRA)$ where all arithmetic atoms are of
form $x \LAOP c$ or $d \LAOP c$, such that $c\in\Int$, $d$ is a (possibly free) constant, $x\in \varset$, and $\LAOP \in \{\leq, <, \neq, =, >, \geq\}$.
\end{definition}

\begin{example}\label{ex_bs_lra_ecu_nonground}
The ECU use case leads to $\BS(\LRA)$ clauses such as
\begin{align}\begin{split}
  &x_1 < y_1 \;\lor\; x_1\geq y_2 \;\lor\; x_2 < y_3  \;\lor\; x_2\geq y_4 \;\lor\; \neg \text{KPa}(x_3, x_1) \;\lor{} \\
  &  \neg \text{Rpm}(x_4, x_2) \;\lor\; \neg \text{IgnTable}(y_1,y_2,y_3,y_4,z)  \;\lor\; \text{IgnDeg1}(x_3, x_4, x_1, x_2, z).
\end{split}\label{eq_bs_lra_ecu_nonground}
\end{align}
This clause is not in $\BS(\SB)$, e.g., since $x_1 > x_5$ is not allowed in $\BS(\SB)$.
However, clause \eqref{eq_bs_lra_ecu} of Example~\ref{ex_bs_lra_ecu} is a $\BS(\SB)$ clause that is an instance of \eqref{eq_bs_lra_ecu_nonground},
obtained by the substitution $\{y_1\mapsto 0,y_2\mapsto 13,y_3\mapsto 880,y_4\mapsto 1100\}$. This grounding will eventually be obtained
by resolution on the $\text{IgnTable}$ predicate, because it occurs only positively in ground unit facts.
\end{example}

Example~\ref{ex_bs_lra_ecu_nonground} shows that $\BS(\SB)$ clauses can sometimes be obtained by instantiation.
Relevant instantiations can be found by \emph{resolution}, in our case by
\emph{hierarchic resolution}, which supports arithmetic constraints:
given clauses $\Lambda_1 \parallel L \lor C_1$ and $\Lambda_2 \parallel K \lor C_2$ with $\sigma = \mMGU(L,\comp(K))$,
their \emph{hierarchic resolvent} is $(\Lambda_1,\Lambda_2 \parallel C_1\lor C_2)\sigma$.
A \emph{refutation} is the sequence of resolution steps that produces a
clause $\Lambda \parallel \bot$ with $\sigval^{\LRA} \models \Lambda\delta$ for some grounding $\delta$.
\emph{Hierarchic resolution} is sound and refutationally complete for pure $\BS(\LRA)$, 
since every set $N$ of pure $\BS(\LRA)$ clauses $N$ is \emph{sufficiently complete}~\cite{BachmairGanzingerEtAl94}, and hence
 \emph{hierarchic superposition} is sound and refutationally complete for $N$~\cite{BachmairGanzingerEtAl94,BaumgartnerWaldmann19}.
Resolution can be used to eliminate predicates that do not occur recursively:

\begin{definition}[Positively Grounded Predicate]
  Let $N$ be a set of $\BS(\LRA)$ clauses. A free first-order predicate $P$ is
  a \emph{positively grounded predicate} in $N$ if all positive occurrences of
  $P$ in $N$ are in ground unit clauses (also called facts).
\end{definition}

For a positively grounded predicate $P$ in a clause set $N$,
let $\elim(P,N)$ be the clause set obtained from $N$ by resolving away
all negative occurrences of $P$ in $N$ and finally eliminating all clauses where $P$ occurs negatively.
We need to keep the $P$ facts for the generation of test points.
Then $N$ is satisfiable iff $\elim(P,N)$ is satisfiable.
We can extend $\elim$ to sets of positively grounded predicates in the obvious way.
If $n$ is the number of $P$ unit clauses in $N$, $m$ the maximal number
of negative $P$ literals in a clause in $N$, and $k$ the number of clauses
in $N$ with a negative $P$ literal, then $|\elim(P,N)| \leq |N| + k\cdot n^m$, i.e.,
$\elim(P,N)$ is exponential in the worst case.

We further assume that $\elim$ simplifies $\LRA$ atoms until they contain at most one integer number and
that $\LRA$ atoms that can be evaluated are reduced to true and false and the respective clause simplified.
For example, given the pure and abstracted $\BS(\LRA)$ clause set
$N = \{\text{IgnTable}(0,13,880,1100,2200),\; x_1\leq x_2 \; \lor\;  z_2 \geq z_1 \parallel  \neg \text{IgnTable}(x_1,x_2,y_1,y_2,z_1)  \;\lor\; \text{R}(z_2)\}$,
the predicate $\text{IgnTable}$ is positively grounded. 
Then $\elim(\text{IgnTable}, N) = \{z_2 \geq 2200 \parallel \text{R}(z_2)\}$ where the unifier $\sigma = \{x_1\mapsto 0, x_2\mapsto 13, y_1 \mapsto 880, y_2\mapsto 110,
z_1\mapsto 2200\}$ is used to eliminate the literal $\neg \text{IgnTable}(x_1,x_2,y_1,y_2,z_1)$ and $(x_1\leq x_2)\sigma$ becomes true and can be removed.

\begin{definition}[Positively Grounded $\BS(\SB)$: $\BS(\SB)\SP$]
  A clause set $N$ is out of the fragment \emph{positively grounded $\BS(\SB)$, $\BS(\SB)\SP$} if
  $\elim(S,N)$ is out of the $\BS(\SB)$ fragment, where $S$ is the set of all
  positively grounded predicates in $N$. 
\end{definition}

Pure $\BS(\SB)\SP$ clause sets are called $\BS(\SB)\PP$ and are the starting point for our Datalog hammer.

\begin{extendedonly}
\begin{lemma}\label{lem:interpretationfrompositivelygrounded}
Let $\inta$ be an interpretation satisfying the clause set $\elim(S,N)$. 
Then we can construct a satisfying interpretation $\inta'$ for $N$ such that $P^{\inta'} = \left\{\bar{a} \in \Real^n \mid P(\bar{a}) \in N \right\}$ if $P \in S$ and otherwise $P^{\inta'} = P^{\inta}$.
\end{lemma}
\begin{prooftoggle}
By contradiction. Assume that $\inta' \nvDash N$. 
Then there must exist a clause $(\Lambda \parallel C) \in N$ and a grounding $\tau: X \rightarrow \Real$ such that $\inta' \nvDash (\Lambda \parallel C)\tau$.
We can split the clause $C = D \vee D'$ into two clauses $D$ and $D'$ such that $D$ contains all literals $\neg P(\bar{t})$ from $C$ with $P \in S$. 
The clause $D'$ does not contain any positive literals $P(\bar{t})$ with $P \in S$ or else $\Lambda \parallel C$ would simplify to a fact $P(\bar{a}) \in N$ that is satisfied by $\inta'$.
Since $P^{\inta'} = \left\{\bar{a} \in \Real^n \mid P(\bar{a}) \in N \right\}$ for $P \in S$, 
we can also assume that any literal $\neg P(\bar{a})$ in $D\sigma$ must correspond to a fact $P(\bar{a}) \in N$ or $D$ would be satisfied by $P^{\inta'}$. 
This set of facts can be defined as $S' = \{P(\bar{a}) \mid \neg P(\bar{a}) \in D\sigma\}$. 
As a result, there exists a clause $(\Lambda' \parallel C') \in \elim(S,N)$ such that 
$(\Lambda' \parallel C')$ is the result of resolving $(\Lambda \parallel D')$ with $S'$; which also means that $(\Lambda' \parallel C')\tau$ is equivalent to $(\Lambda \parallel D')\tau$. 
Moreover, $\inta' \vDash (\Lambda' \parallel C') \cdot\tau$ because $\inta'$ behaves the same as $\inta$ on all clauses without any literal over a $P \in S$. 
Hence, $\inta' \vDash (\Lambda \parallel C)\tau$ which is a contradiction to our initial assumption, so $\inta' \vDash N$.
\end{prooftoggle}

\begin{lemma}\label{lem:interpretationforpositivelygrounded}
Every interpretation $\inta$ that satisfies the clause set $N$ also satisfies $\elim(S, N)$.
\end{lemma}
\begin{prooftoggle}
  By soundness of hierarchic resolution.
\end{prooftoggle}
\end{extendedonly}


\section{The Theory of the Hammer} \label{sec:theory}

We define two hammers that help us solve $\BS(\SB)\PP$ clause sets with both universally and existentially quantified conjectures.
Both are equisatisfiability preserving and allow us to abstract $\BS(\SB)\PP$ formulas into less complicated logics with efficient and complete decision procedures.

The first hammer, also called \emph{grounding hammer}, translates any $\BS(\SB)\PP$ clause set $N$ with a universally/existentially quantified conjecture into an equisatisfiable ground and no longer pure $\BS(\SB)$ clause set over a finite
set of first-order constants called \emph{test points}.
This means we reduce a quantified problem over an infinite domain into a ground problem over a finite domain. The size of the ground problem
grows worst-case exponentially in the number of variables and the number of numeric constants in $N$ and the conjecture.
For the Horn case, $\HBS(\SB)\PP$, we define a Datalog hammer, i.e. a transformation into an equisatisfiable Datalog program that is
based on the same set of test points but does not require an overall grounding.
It keeps the original clauses almost one-to-one instead of greedily computing all ground instances of those clauses over the test points.
The Datalog hammer adds instead a finite set of Datalog facts that correspond to all theory atoms over the given set of test points.
With the help of these facts and the original rules, the Datalog reasoner can then derive the same conclusions as it could have done with the ground \HBS(\SB) clause set, however,
all groundings that do not lead to new ground facts are neglected.
Therefore, the Datalog approach is much faster in practice because the Datalog reasoner wastes no time (and space)
on trivially satisfied ground rules that would have been part of the greedily computed ground \HBS(\SB) clause set.
Moreover, Datalog reasoners are well suited to the resulting structure of the problem, i.e. many facts but a small set of rules.

Note that we never compute or work on $\elim(S, N)$ although the discussed clause sets are positively grounded.
We only refer to $\elim(S, N)$ because it allows us to formulate our theoretical results more concisely.
We avoid working on $\elim(S, N)$ because it often increases the number of non-fact clauses (by orders of magnitude) in order to simplify the positively grounded theory atoms to variable bounds.
This is bad in practice because the number of non-fact clauses has a high impact on the performance of Datalog reasoners.
Our Datalog hammer resolves this problem by dealing with the positively grounded theory atoms in a different way that only introduces more facts instead of non-fact clauses.
This is better in practice because Datalog reasoners are well suited to handling a large number of facts.
Since the \emph{grounding hammer} is meant primarily as a stepping stone towards the Datalog hammer, we also defined it in such a way that it avoids computing and working on $\elim(S, N)$.

\subsubsection{Hammering $\BS(\SB)$ Clause Sets with a Universal Conjecture:}
Our first hammer, takes a $\BS(\SB)\PP$ clause set $N$ and a universal conjecture $\forall \bar{y}.\phi$ as input and translates it into a ground $\BS(\SB)$ formula.
We will later show that the cases for no conjecture and for an existential conjecture can be seen as special cases of the universal conjecture.
Since $\phi$ is a universal conjecture, we assume that $\phi$ is a quantifier-free pure $\BS(\SB)$ formula and $\vars(\phi) = \vars(\bar{y})$.
Moreover, we denote by $S$ the set of positively grounded predicates in $N$ 
and assume that none of the positively grounded predicates from $S$ appear in $\phi$.
There is not much difference developing the hammer for the Horn or
the non-Horn case. Therefore, we present it
for the general non-Horn case, although our second Datalog hammer is restricted to Horn.
Note that a conjecture $\forall \bar{y}.\phi$ is a consequence of $N$, i.e. 
$N \models \forall \bar{y}.\phi$, if $\forall \bar{y}.\phi$ is satisfied by every interpretation $\sigval$ that also satisfies $N$, i.e. $\forall \sigval. (\sigval \models N \rightarrow \forall \bar{y}.\phi)$.
Conversely, $\forall \bar{y}.\phi$ is not a consequence of $N$ if there exists a counter example, i.e. one interpretation $\sigval$ that satisfies $N$ but does not satisfy $\forall \bar{y}.\phi$, or formally: $\exists \sigval. (\sigval \models N \wedge \exists \bar{y}. \neg \phi)$. 

Our hammer is going to abstract the counter example formulation into a ground $\BS(\SB)$ formula.
This means the hammered formula will be unsatisfiable if and only if the conjecture is a consequence of $N$.
The abstraction to the ground case works because we can restrict our solution space from the infinite reals to a finite set of test points and still preserve satisfiability. 
To be more precise, we partition $\Real$ into intervals such that any variable bound in $\elim(S, N)$ and $\phi$ either satisfies all points in one such interval $I$ or none. 
Then we pick $m = \max(1,|\vars(\phi)|)$ test points from each of those intervals because 
any counter example, i.e. any assignment for $\neg \phi$, contains at most $m$ different points per interval.

We get the interval partitioning by first determining the necessary set of interval borders based on the variable bounds in $\elim(S, N)$ and $\phi$. 
Then, we sort and combine the borders into actual intervals.
The interval borders are extracted as follows: 
We turn every variable bound $x \LAOP c$ with $\LAOP \in \{\leq,<,>,\geq\}$ in $\elim(S, N)$ and $\phi$ into two interval borders. One of them is the interval border implied by the bound itself and the other its negation, e.g., $x \geq 5$ results in the interval border \dblquotes{}$[5$\dblquotes{} and the interval border of the negation \dblquotes{}$5)$\dblquotes{}.
Likewise, we turn every variable bound $x \LAOP c$ with $\LAOP \in \{=,\neq\}$ into all four possible interval borders for $c$, i.e. \dblquotes{}$c)$\dblquotes{}, \dblquotes{}$[c$\dblquotes{}, \dblquotes{}$c]$\dblquotes{}, and \dblquotes{}$(c$\dblquotes{}.
The set of interval endpoints $\mathcal{C}$ is then defined as follows:
\begin{notinextended}
\[\begin{array}{l l}
\mathcal{C} = &\left\{\dblquotes{}c]\dblquotes{}, \dblquotes{}(c\dblquotes{} \mid x \triangleleft c \in \atoms(\elim(S, N))\cup\atoms(\phi) \text{ where } \triangleleft \in \{\leq,=,\neq,>\} \right\} \; \cup \\
 &\left\{\dblquotes{}c)\dblquotes{}, \dblquotes{}[c\dblquotes{} \mid x \triangleleft c \in \atoms(\elim(S, N))\cup\atoms(\phi)  \text{ where } \triangleleft \in \{\geq,=,\neq,<\} \right\} \; \cup \; \{\dblquotes{}(-\infty\dblquotes{},\dblquotes{}\infty)\dblquotes{}\}
\end{array}\]
\end{notinextended}
\begin{extendedonly}
\[\begin{array}{l l}
\mathcal{C} = &\left\{\dblquotes{}c]\dblquotes{}, \dblquotes{}(c\dblquotes{} \mid x \triangleleft c \in \atoms(\elim(S, N))\cup\atoms(\phi) \text{ where } \triangleleft \in \{\leq,=,\neq,>\} \right\} \; \cup \\
 &\left\{\dblquotes{}c)\dblquotes{}, \dblquotes{}[c\dblquotes{} \mid x \triangleleft c \in \atoms(\elim(S, N))\cup\atoms(\phi)  \text{ where } \triangleleft \in \{\geq,=,\neq,<\} \right\} \; \cup \\ &\{\dblquotes{}(-\infty\dblquotes{},\dblquotes{}\infty)\dblquotes{}\}
\end{array}\]
\end{extendedonly}
It is not necessary to compute $\elim(S, N)$ to compute $\mathcal{C}$. 
It is enough to iterate over all theory atoms in $N$ and compute all of their instantiations in $\elim(S, N)$ based on the facts in $N$ for predicates in $S$. 
This can be done in $O(n_t \cdot n_A \cdot n_S^{n_v})$, where $n_v$ is the maximum number of variables in any theory atom in $N$,
$n_A$ is the number of theory atoms in $N$, $n_S$ is the number of facts in $N$ for predicates in $S$, and $n_t$ is the size of the
largest theory atom in $N$ with respect to the number of symbols.

The intervals themselves can be constructed by sorting $\mathcal{C}$ in an ascending order such that 
we first order by the border value---i.e. $\delta < \epsilon$ if $\delta \in \{\dblquotes{}c)\dblquotes{}, \dblquotes{}[c\dblquotes{}, \dblquotes{}c]\dblquotes{}, \dblquotes{}(c\dblquotes{}\}$, $\epsilon \in \{\dblquotes{}d)\dblquotes{}, \dblquotes{}[d\dblquotes{}, \dblquotes{}d]\dblquotes{}, \dblquotes{}(d\dblquotes{}\}$, and $c < d$---and 
then by the border type---i.e. $\dblquotes{}c)\dblquotes{} < \dblquotes{}[c\dblquotes{} < \dblquotes{}c]\dblquotes{} < \dblquotes{}(c\dblquotes{}$.
The result is a sequence $[\ldots,\dblquotes{}\delta_l\dblquotes{},\dblquotes{}\delta_u\dblquotes{},\ldots]$, 
where we always have one lower border $\delta_l$, 
followed by one upper border $\delta_u$. 
We can guarantee that an upper border $\delta_u$ follows a lower border $\delta_l$ because $\mathcal{C}$ always contains \dblquotes{}$c)$\dblquotes{} together with \dblquotes{}$[c$\dblquotes{} and \dblquotes{}$c]$\dblquotes{} together with \dblquotes{}$(c$\dblquotes{} for $c \in \Int$, so 
always two consecutive upper and lower borders.
Together with \dblquotes{}$(-\infty$\dblquotes{} and \dblquotes{}$\infty)$\dblquotes{} this guarantees that the sorted $\mathcal{C}$ has the desired structure.
If we combine every two subsequent borders $\delta_l$, $\delta_u$ in our sorted sequence $[\ldots,\dblquotes{}\delta_l\dblquotes{},\dblquotes{}\delta_u\dblquotes{},\ldots]$, then we receive our partition of intervals $\intvp$.
For instance, if $x < 5$ and $x = 0$ are the only variable bounds in $\elim(S, N)$ and $\phi$, then $\mathcal{C} = \{\dblquotes{}5)\dblquotes{},\dblquotes{}[5\dblquotes{},\dblquotes{}0)\dblquotes{},\dblquotes{}[0\dblquotes{},\dblquotes{}0]\dblquotes{},\dblquotes{}(0\dblquotes{},\dblquotes{}(-\infty\dblquotes{},\dblquotes{}\infty)\dblquotes{}\}$ and if we sort it we get $\{(-\infty,0),[0,0],(0,5),[5,\infty)\}$.

\begin{corollary}\label{cor:uniformtheory}
Let  $\triangleleft \in \{<,\leq,=,\neq,\geq,>\}$. For each interval $I \in \intvp$, every two points $a, b \in I$, and every variable bound $x \triangleleft c \in \atoms(\elim(S, N))\cup\atoms(\phi)$, $a \triangleleft c$ if and only if $b \triangleleft c$.
\end{corollary}

The above Corollary states that two points $a, b \in I$ belonging to the same interval $I \in \intvp$ satisfy the same theory atoms in $\elim(S, N)$ and $\phi$.
However, two points $a, b \in I$ do not necessarily satisfy the same 
non-theory atom under an arbitrary interpretation $\sigval$;
not even if $\sigval$ satisfies $N \wedge \exists \bar{y}. \neg \phi$.
E.g., $\sigval$ may evaluate $P(a)$ to true and $P(b)$ to false.
Sometimes this is even necessary or we would be unable to find a counter example:

\begin{example}
Let $\phi = (0 \leq x, x \leq 1, 0 \leq y, y \leq 1 || \neg P(x) \vee P(y))$ be our conjecture and $N = \emptyset$ be our clause set.
Informally, the property $\forall x,y.\phi$ states that $P$ must be uniform over the interval $[0,1]$, i.e. either all points in the interval $[0,1]$ satisfy $P$ or none do.
As a result, all interpretations that are uniform over $[0,1] \in \intvp$ also satisfy $\forall x,y.\phi$. 
However, there still exist counter examples that are not uniform, e.g., $P^\sigval = \{0\}$, which satisfies $N$ but not $\forall x,y.\phi$ because it evaluates $P(0)$ to true and $P(a)$ to false for all $a \in [0,1] \setminus \{0\}$.
\end{example}

To better understand the above example, let us look again at the counter example formulation $N \wedge \exists \bar{y}. \neg \phi$.
This formula is satisfiable, i.e. we have a counter example to our conjecture $\forall \bar{y}. \phi$ 
if there exists an interpretation $\sigval$ and a grounding $\rho$ for $\phi$ (also called an assignment for $\phi$) such that $\sigval$ satisfies $N$ and $\neg \phi \rho$.
In the worst case, the assignment $\rho$ maps to $m=|\vars(\phi)|$ different points in one of the intervals $I \in \intvp$.
Each of those $m$ points may "act" differently in the interpretation $\sigval$ although it belongs to the same interval.
On the one hand, this means that we need in the worst case $m=|\vars(\phi)|$ different test points for each interval in $\intvp$.
On the other hand, we will show in the proof of Lemma~\ref{lem:verificationsampling} that we can always find a counter example, where (i)~no more than $m$ points per interval act differently and (ii)~the actual value of a point does not matter as long as it belongs to the same interval $I \in \intvp$.
This is owed mainly to Corollary~\ref{cor:uniformtheory}, i.e. that the points in an interval act at least the same in the theory atoms.
We ensure that a test point $a$ belongs to a certain interval $I$ by adding a set of variable bounds to our formula. We define these bounds with the functions $\ilbnd$ and $\iubnd$ that turn intervals into lower and upper bounds: $\ilbnd((-\infty,u),x) = \emptyset$, $\ilbnd((-\infty,u],x) = \emptyset$, $\ilbnd((l,u),x) = \{l < x\}$, $\ilbnd((l,u],x) = \{l < x\}$, $\ilbnd([l,u),x) = \{l \leq x\}$, $\ilbnd([l,u],x) = \{l \leq x\}$ for $l \neq - \infty$; $\iubnd((l,\infty),x) = \emptyset$, $\iubnd([l,\infty),x) = \emptyset$, $\iubnd((l,u),x) = \{x < u\}$, $\iubnd((l,u],x) = \{x \leq u\}$, $\iubnd([l,u),x) = \{x < u\}$, $\iubnd([l,u],x) = \{x \leq u\}$ for $u \neq \infty$.

Note that this test point scheme would no longer be possible if we were to allow general inequalities.
Even allowing difference constraints, i.e., inequalities of the form $x - y \leq c$, 
would turn the search for a counter example into an undecidable problem~\cite{Downey1972,HorbachEtAl17ARXIV},
because variables can now interact both on the first-order and the theory side.

As a result of these observations, we construct the hammered formula $\psi$, also
called the \emph{finite abstraction} of $N \wedge \exists \bar{y}. \neg \phi$, as follows.
First we fix the following notations for the remaining subsection: 
$\intvp$ is the interval partition for $N$ and $\phi$;
$\intvp_{=}=\{I \in \intvp \mid I = [l,l] \}$ is the set of all intervals from $\intvp$ that are just points; 
$\intvp_{\infty}=\intvp \setminus \intvp_{=}$ is the set of all intervals that are not just points and therefore contain infinitely many values;
$m = \max(1,|\vars(\phi)|)$ is the number of test points needed per interval with infinitely many values;
$B = \{a_{I,1} | I \in \intvp_{=}\} \cup \{a_{I,j} | I \in \intvp_{\infty} \text{ and } j = 1,\ldots,m\}$ is the set of test points for our abstraction such that we have one test point per interval $I\in \intvp_{=}$ and $m$ different test points for each interval $I \in \intvp_{\infty}$;
$\idef(B) = \bigcup_{a_{I,i} \in B} \ilbnd(I,a_{I,i}) \cup \bigcup_{a_{I,i} \in B} \iubnd(I,a_{I,i})$ is a set of bounds that defines to which interval each constant belongs;
 and 
$\psi= \mGnd_B(N) \cup \idef(B) \wedge ( \bigvee_{\rho : \vars(\phi) \rightarrow B} \neg \phi  \rho )$ is the finite abstraction of $N \wedge \exists \bar{y}. \neg \phi$.

The hammered formula $\psi$ contains $\mGnd_B(N)$, i.e. a ground clause $(\Lambda \parallel C) \sigma$ for every clause $(\Lambda \parallel C) \in N$ and every assignment $\sigma: \vars(\Lambda \parallel C) \rightarrow B$.
This means any deduction over the tests points $B$ we could have performed with the set of clauses $N$ can also be performed with the set of clauses $\mGnd_B(N)$ in $\psi$.
Similarly, $\bigvee_{\rho : \vars(\phi) \rightarrow B} \neg \phi \rho$ is a big disjunction over all assignments of $\rho$ for $\phi$ that assign its variables to test points.
Hence, $\psi$ is satisfiable if there exists a counter example for $N \wedge \exists \bar{y}. \neg \phi$ that just uses the test points $B$.
Although the finite abstraction is restricted to the test points $B$, it is easy to extend any of its interpretations to all of $\Real$ and our original formula.
We just have to interpret all values in an interval that are not test points like one of the test points:

\begin{lemma}\label{lem:verificationextrapolation}
Let $\sigval'$ be an interpretation satisfying the finite abstraction $\psi$ of $N \wedge \exists \bar{y}. \neg \phi$.
Moreover, let $\rho : \vars(\phi) \rightarrow B$ be a substitution such that $\sigval'$ satisfies $\neg \phi \rho$.
Then the interpretation $\sigval$ satisfies $N \wedge \exists \bar{y}. \neg \phi$ if it is constructed as follows:\newline
$P^{\sigval} = \{\bar{a} \in \Real^n \mid P(\bar{a}) \in N\}$ if $P \in S$ and $P^{\sigval} =  \{\bar{a} \in \Real^n \mid \bar{a}  \sigma \in P^{\sigval'}\}$ if $P \not\in S$ and $\sigma = \{a \mapsto a_{I,1}^{\sigval'} \mid I \in \intvp \text{ and } a \in I \setminus \{a_{I,2}^{\sigval'} , \ldots, a_{I,m}^{\sigval'}\}\}$.
\end{lemma}
\begin{prooftoggle}
Before we start with the actual proof,
we need to define a second substitution $\sigma' = \{a \mapsto a_{I,1} \mid I \in \intvp \text{ and } a \in I \setminus \{a_{I,2}^{\sigval'} , \ldots, a_{I,m}^{\sigval'}\}\} \cup \{a \mapsto a_{I,j} \mid a = a_{I,j}^{\sigval'}\}$ 
that maps any point $a$ in one of our intervals $I$, 
either to the test point $a_{I,j}$ that is interpreted by $\sigval'$ as $a$ (i.e. $a = a_{I,j}^{\sigval'}$) or to the default test point for the interval $a_{I,1}$.
Then we split the proof into two parts: 
1)~we show that $\sigval$ satisfies $N$; 
2)~after that we show that there exists a $\rho'$ such that $\sigval$ satisfies $\neg \phi \rho'$.\newline
1) Instead of directly showing that $\sigval$ satisfies $N$, 
we show that $\sigval$ satisfies $\elim(S, N)$.
Thanks to Lemma~\ref{lem:interpretationfrompositivelygrounded}, 
we then also know that $\sigval$ satisfies $N$. 
Note that $\sigval$ satisfies $\elim(S, N)$ is equivalent to
$\sigval$ satisfies $(\Lambda \parallel C) \tau$ for all $(\Lambda \parallel C) \in \elim(S, N)$ and all groundings $\tau: \vars(\Lambda \parallel C) \rightarrow \Real$.
However, for any $(\Lambda \parallel C) \in \elim(S, N)$ and any grounding $\tau: \vars(\Lambda \parallel C) \rightarrow \Real$, 
we can construct an alternative grounding to $B$ instead of $\Real$\newline
\centerline{$\tau' = \left\{x \mapsto (x \tau \sigma') \mid x \in \vars(\Lambda \parallel C) \right\}$}
that applies $\sigma'$ to the result values of $\tau$, which means we map the result values to their corresponding values in our set of test points $B$.
We know that $\sigval' \models (\Lambda \parallel C) \tau'$ because $\sigval' \models \psi$ and there exists a $(\Lambda' \parallel C') \in \mGnd_B(N)$ and a set of ground facts $S' \in \mGnd_B(N)$ that resolve to $(\Lambda \parallel C)  \tau'$ by definition of $\elim(S, N)$.
Due to the definition of $\sigval$, we also know that $\bar{a} \in P^{\sigval}$ if and only if $\bar{a} \sigma \in P^{\sigval'}$.
Similarly, Corollary~\ref{cor:uniformtheory} and the definition of the $a_{I,i} \in B$ imply that $a \in \mathbb{R}$ satisfies a variable bound from $\elim(S, N)$ if and only if $a \sigma$ satisfies it.
Therefore, $\sigval \models (\Lambda \parallel C) \tau$ if and only if $\sigval' \models (\Lambda \parallel C) \tau'$.
Hence, $\sigval$ satisfies $\elim(S, N)$ and by Lemma~\ref{lem:interpretationfrompositivelygrounded} also $N$.\newline
2) We start by constructing a substitution from the variables of $\phi$ to $\Real$: $\rho' = \{y_i \mapsto a_{I,j}^{\sigval'} \mid y_i \rho = a_{I,j}\}$.
By definition, $\sigval$ satisfies $\neg \phi  \rho'$ because $\sigval'$ satisfies $\neg \phi  \rho$.\newline
Now the two subproofs combined prove that $\sigval$ is satisfying $N \wedge \exists \bar{y}. \neg \phi$.
\end{prooftoggle}

Similarly, we can extend any interpretation $\sigval$ satisfying $N \wedge \exists \bar{y}. \neg \phi$ into an interpretation satisfying $\psi$.
We just have to pick one assignment $\rho': \vars(\phi) \rightarrow \Real$ such that 
$\sigval$ satisfies $\neg \phi \rho'$ and pick one test point $B$ for each point in $\cdom(\rho')$ and interpret it as its corresponding point in $\cdom(\rho')$.

\begin{lemma}\label{lem:verificationsampling}
Let $\sigval$ be an interpretation satisfying the formula $N \wedge \exists \bar{y}. \neg \phi$.
Then we can construct an interpretation $\sigval'$ that satisfies its finite abstraction $\psi$.
\end{lemma}
\begin{prooftoggle}
If $\sigval$ satisfies $N \wedge \exists \bar{y}. \neg \phi$, then 
$\sigval$ satisfies $N$ and 
there exists a $\rho': \vars(\phi) \rightarrow \Real$ such that 
$\sigval$ also satisfies $\neg \phi  \rho'$.
We extend the interpretation $\sigval$ so it also satisfies $\psi$ 
by interpreting some of our test points $B$
as points in $B' = \cdom(\rho')$
and all other test points as random values $a_I \in I \setminus B'$ belonging to their corresponding interval $I$.
Explicitly this means that we extend the interpretation $\sigval$ to our constants $B$ as follows:\newline
$a_{I,j}^{\sigval} := b_j$ if $j \leq |B' \cap I|$ and $B' \cap I = \{b_1,\ldots,b_k\}$;
$a_{I,j}^{\sigval} := a_I$ if $j > |B' \cap I|$.\newline
This means if $k$ points $\{b_1,\ldots,b_k\}$ in the codomain $B'$ of $\rho'$ belong to the interval $I$,
then the first $k$ test points $a_{I,1}, \ldots, a_{I,k}$ are interpreted as $b_1,\ldots,b_k$ and all other test points $a_{I,j}$ as $a_I$.
As a result, the extended interpretation $\sigval$ will have one test point $a_{I,j} \in B$ for every $b \in B'$ that will be interpreted by $\sigval$ as $b$.
Next, we construct an assignment $\rho^*$ for $\phi$ that ranges into $B$ such that $\sigval$ satisfies $\neg \phi \rho^*$. The assignment $\rho^* := \{x \mapsto a_{I,j} \mid x \in \vars(\phi) \text{ and } x \rho' = a_{I,j}^{\sigval}\}$
is almost the same as our assignment $\rho'$ that falsified our conjecture over $\Real$, 
except that the result values of $\rho'$ are swapped with their corresponding test points in $B$.
By definition of the extended $\sigval$, $\sigval$ satisfies $\mGnd_B(N)$ because $\sigval$ satisfies $N$.
Due to the way we extended $\sigval$ over the constants $B$, $\sigval$ also satisfies each bound in $\idef(B)$.
By definition of $\rho^*$ and $\rho'$, $\sigval$ also satisfies $\neg \phi \rho^*$ and therefore $\left( \bigvee_{\rho : \vars(\phi) \rightarrow B} \neg \phi  \rho \right)$. 
Hence, the extended $\sigval$ satisfies $\psi$.
\end{prooftoggle}

If we combine both results, we get that $N \wedge \exists \bar{y}. \neg \phi$ is equisatisfiable to $\psi$:

\begin{lemma}\label{lem:verificationequiv}
$N \wedge \exists \bar{y}. \neg \phi$ has a satisfying interpretation if and only if its finite abstraction $\psi$ has a satisfying interpretation. 
\end{lemma}
\begin{prooftoggle}
The first part of the equivalence follows from Lemma~\ref{lem:verificationextrapolation}.
The second part follows from Lemma~\ref{lem:verificationsampling}.
\end{prooftoggle}

The finite abstraction for the case with a universal conjecture can also be used to construct a finite abstraction for the case without a conjecture and the case with an existential conjecture.
Let $N$ be a $\BS(\SB)\PP$ clause set and let $S$ be the set of all positively grounded predicates in $N$. 
$N$ is satisfiable if and only if $N \not\models \bot$. 
Hence, we get a finite abstraction for $N$ if we build one for $N \models \bot$, which can be treated as a universal conjecture because all variables in $\bot$ are universally quantified.
The existential case works similarly: $N \models \exists \bar{y}.\phi$ if and only if $N \cup N' \models \bot$, where $N'$ is the universal $\BS(\SB)$ clause set we get from applying a CNF transformation~\cite{NonnengartW01} to $\forall \bar{y}.\neg \phi$.

\begin{extendedonly}
\begin{example}
We finish the presentation of our first hammer by applying it to two examples $N \models \forall x, y.\phi_1$ and $N \models \forall x, y.\phi_2$. 
For the examples we choose\newline
\centerline{$N := \{0 \leq x, x \leq 2 \parallel \neg P(x) \vee Q(x), \quad x \leq 1 \parallel P(x), \quad x > 1 \parallel \neg P(x)\}$}
as our set of clauses and check two different conjectures \newline
\centerline{$\phi_1 := 0 \leq x, x \leq 1, 0 \leq y, y \leq 1 \parallel \neg Q(x) \vee Q(y)$}
and\newline
\centerline{$\phi_2 := 1 < x, x \leq 2, 1 < y, y \leq 2 \parallel \neg Q(x) \vee Q(y)$.}
The conjecture $\phi_1$ is true if $Q$ is always interpreted uniformly over the interval $[0,1]$, i.e. for every interpretation $\sigval$ satisfying $N$ either $\sigval$ evaluates $Q$ to true for all points in $[0,1]$ ($Q^{\sigval} = [0,1]$) or to false for all points in $[0,1]$ ($Q^{\sigval} = \emptyset$).
The conjecture $\phi_2$ states almost the same property except that $Q$ has to be interpreted uniformly over the interval $(1,2]$.
We now construct for both examples their finite abstraction. 
To this end, we first compute their sets of interval endpoints,
which are both equivalent because $N$ has no positively grounded predicates and all variable bounds in our conjectures $\phi_1$ and $\phi_2$ also appear in $N$.
The variable bounds in $N$ are 
$0 \leq x$, $x \leq 2$, $x \leq 1$, $x > 1$.
As a result, we get as the set of interval endpoints $\mathcal{C} = \{\dblquotes{}0)\dblquotes{},\dblquotes{}[0\dblquotes{},\dblquotes{}2]\dblquotes{},\dblquotes{}(2\dblquotes{},\dblquotes{}1]\dblquotes{},\dblquotes{}(1\dblquotes{},\dblquotes{}(-\infty\dblquotes{},\dblquotes{}\infty)\dblquotes{}\}$.
Next we sort an recombine the endpoints in $\mathcal{C}$,
and receive the interval partition $\intvp = \{(-\infty,0),[0,1],(1,2],(2,\infty)\}$.
Since our conjectures contain two variables, we need two test points/constants for each interval.
Therefore, $B = \{a_{(-\infty,0),1}, a_{(-\infty,0),2}, a_{[0,1],1}, a_{[0,1],2}, a_{(1,2],1}, a_{(1,2],2}, a_{(2,\infty),1}, a_{(2,\infty),2}\}$.
For both finite abstractions $\psi_1 = \mGnd_B(N) \cup \idef(B) \wedge ( \bigvee_{\rho : \vars(\phi_1) \rightarrow B} \neg \phi_1  \rho )$ and $\psi_2 = \mGnd_B(N) \cup \idef(B) \wedge ( \bigvee_{\rho : \vars(\phi_2) \rightarrow B} \neg \phi_2  \rho )$,
$\mGnd_B(N)$ and $\idef(B)$ are the same.
As mentioned before, 
$\idef(B)$ formally defines that each $a_{I,j}$ must be interpreted as a value in its respective interval $I$ with the help of variable bounds: \newline
\centerline{$\begin{array}{l l}
\idef(B) := \{ &a_{(-\infty,0),1} < 0, a_{(-\infty,0),2} < 0, 0 \leq a_{[0,1],1}, a_{[0,1],1} \leq 1, 0 \leq a_{[0,1],2},\\
&a_{[0,1],2} \leq 1, 1 < a_{(1,2],1}, a_{(1,2],1} \leq 2, 1 < a_{(1,2],2}, a_{(1,2],2} \leq 2,\\
&2 < a_{(2,\infty),1}, 2 < a_{(2,\infty),2}\}
\end{array}$}
$\mGnd_B(N)$ defines all groundings of the clauses $N$ over the set of test points $B$, i.e.\newline
\centerline{$\begin{array}{l l}
\mGnd_B(N) := \{&0 \leq a_{(-\infty,0),1}, a_{(-\infty,0),1} \leq 2 \parallel \neg P(a_{(-\infty,0),1}) \vee Q(a_{(-\infty,0),1}), \ldots,\\
&0 \leq a_{(2,\infty),2}, a_{(2,\infty),2} \leq 2 \parallel \neg P(a_{(2,\infty),2}) \vee Q(a_{(2,\infty),2}),\\
&a_{(-\infty,0),1} \leq 1 \parallel P(a_{(-\infty,0),1}), \ldots, a_{(2,\infty),2} \leq 1 \parallel P(a_{(2,\infty),2}),\\
&a_{(-\infty,0),1} > 1 \parallel \neg P(a_{(-\infty,0),1}), \ldots, 
a_{(2,\infty),2} > 1 \parallel \neg P(a_{(2,\infty),2})
\}
\end{array}$}
Note that half of the clauses $(\Lambda \parallel C) \in \mGnd_B(N)$ are trivially satisfied because the inequalities in $\idef(B)$ ensure that $\Lambda$ is not satisfiable, e.g. $a_{(-\infty,0),1} > 1 \parallel \neg P(a_{(-\infty,0),1})$ is trivially satisfied because $a_{(-\infty,0),1} > 1$ and $(a_{(-\infty,0),1} < 0) \in \idef(B)$ contradict each other.
As a result, we can remove those clauses without loss of generality.
Similarly, all other clauses $(\Lambda \parallel C) \in \mGnd_B(N)$ satisfy their theory atoms $\Lambda$ if we take the inequalities in $\idef(B)$ into account, e.g. in $0 \leq a_{(1,2],1}, a_{(1,2],1} \leq 2 \parallel \neg P(a_{(1,2],1}) \vee Q(a_{(1,2],1})$ the inequalities $(1 < a_{(1,2],1}) \in \idef(B)$ and $(a_{(1,2],1} \leq 2) \in \idef(B)$ ensure that $0 \leq a_{(1,2],1}$ and $a_{(1,2],1} \leq 2$.
Therefore, we can remove their theory atoms without loss of generality, i.e. simplify $\Lambda \parallel C$ into $C$.
So at least for our intuitive understanding, we can simplify $\mGnd_B(N)$ to\newline
\centerline{$\begin{array}{ll}
N_G = \{
&\neg P(a_{[0,1],1}) \vee Q(a_{[0,1],1}), \quad
\neg P(a_{[0,1],2}) \vee Q(a_{[0,1],2}), \;\\
&\neg P(a_{(1,2],1}) \vee Q(a_{(1,2],1}), \quad
\neg P(a_{(1,2],2}) \vee Q(a_{(1,2],2}), \;\\
&P(a_{(-\infty,0),1}), \quad
P(a_{(-\infty,0),2}), \quad
P(a_{[0,1],1}), \quad
P(a_{[0,1],2}), \;\\
&\neg P(a_{(1,2],1}), \quad
\neg P(a_{(1,2],2}), \quad
\neg P(a_{(2,\infty),1}), \quad
\neg P(a_{(2,\infty),2})\}
\end{array}$}
The finite abstractions do, however, differ in the groundings of their two conjectures:
\centerline{$\begin{array}{l}
\bigvee_{\rho : \vars(\phi_1) \rightarrow B} \neg \phi_1  \rho := \phi^*_1 := \\
\neg \left(0 \leq a_{(-\infty,0),1}, a_{(-\infty,0),1} \leq 1, 0 \leq a_{(-\infty,0),1}, a_{(-\infty,0),1} \leq 1 \parallel \neg Q(a_{(-\infty,0),1}) \vee Q(a_{(-\infty,0),1})\right)
\; \vee \\
\neg \left(0 \leq a_{(-\infty,0),1}, a_{(-\infty,0),1} \leq 1, 0 \leq a_{(-\infty,0),2}, a_{(-\infty,0),2} \leq 1 \parallel \neg Q(a_{(-\infty,0),1}) \vee Q(a_{(-\infty,0),2})\right)
\; \vee \\
\neg \left(0 \leq a_{(-\infty,0),1}, a_{(-\infty,0),1} \leq 1, 0 \leq a_{[0,1],1}, a_{[0,1],1} \leq 1 \parallel \neg Q(a_{(-\infty,0),1}) \vee Q(a_{[0,1],1})\right)
\; \vee \\
\ldots \vee 
\neg \left(0 \leq a_{(2,\infty),2}, a_{(2,\infty),2} \leq 1, 0 \leq a_{(2,\infty),2}, a_{(2,\infty),2} \leq 1 \parallel \neg Q(a_{(2,\infty),2}) \vee Q(a_{(2,\infty),2})\right)
\end{array}$}
\centerline{$\begin{array}{l}
\bigvee_{\rho : \vars(\phi_2) \rightarrow B} \neg \phi_2  \rho := \phi^*_2 := \\
\neg \left(1 < a_{(-\infty,0),1}, a_{(-\infty,0),1} \leq 2, 1 < a_{(-\infty,0),1}, a_{(-\infty,0),1} \leq 2 \parallel \neg Q(a_{(-\infty,0),1}) \vee Q(a_{(-\infty,0),1})\right)
\; \vee \\
\neg \left(1 < a_{(-\infty,0),1}, a_{(-\infty,0),1} \leq 2, 1 < a_{(-\infty,0),2}, a_{(-\infty,0),2} \leq 2 \parallel \neg Q(a_{(-\infty,0),1}) \vee Q(a_{(-\infty,0),2})\right)
\; \vee \\
\neg \left(1 < a_{(-\infty,0),1}, a_{(-\infty,0),1} \leq 2, 1 < a_{[0,1],1}, a_{[0,1],1} \leq 2 \parallel \neg Q(a_{(-\infty,0),1}) \vee Q(a_{[0,1],1})\right)
\; \vee \\
\ldots \vee 
\neg \left(1 < a_{(2,\infty),2}, a_{(2,\infty),2} \leq 2, 1 < a_{(2,\infty),2}, a_{(2,\infty),2} \leq 2 \parallel \neg Q(a_{(2,\infty),2}) \vee Q(a_{(2,\infty),2})\right)
\end{array}$}
But similarly to $\mGnd_B(N)$, we can drastically simplify the groundings of our conjectures if we take $\idef(B)$ into account (and some other boolean simplifications).
For instance $\neg (1 < a_{(-\infty,0),1}, a_{(-\infty,0),1} \leq 2, 1 < a_{(-\infty,0),1}, a_{(-\infty,0),1} \leq 2 \parallel \neg Q(a_{(-\infty,0),1}) \vee Q(a_{(-\infty,0),1}))$ simplifies to false because the theory atom $1 < a_{(-\infty,0),1}$ must be false according to 
$\idef(B)$ and therefore $(1 < a_{(-\infty,0),1}, a_{(-\infty,0),1} \leq 2, 1 < a_{(-\infty,0),1}, a_{(-\infty,0),1} \leq 2 \parallel \neg Q(a_{(-\infty,0),1}) \vee Q(a_{(-\infty,0),1}))$ is trivially true.
So at least for our intuitive understanding, we can simplify $\bigvee_{\rho : \vars(\phi_1) \rightarrow B} \neg \phi_1  \rho$ to\newline
\centerline{$\phi'_1 := (Q(a_{[0,1],1}) \wedge \neg Q(a_{[0,1],2})) \vee
(Q(a_{[0,1],2}) \wedge \neg Q(a_{[0,1],1}))$}
and $\bigvee_{\rho : \vars(\phi_2) \rightarrow B} \neg \phi_2  \rho$ to\newline
\centerline{$\phi'_2 := (Q(a_{(1,2],1}) \wedge \neg Q(a_{(1,2],0),2})) \vee
(Q(a_{(1,2],2}) \wedge \neg Q(a_{(1,2],1}))$.}
Now based on these simplifications it is relatively easy to show that
$N \models \forall x, y.\phi_1$ and $N \not\models \forall x, y.\phi_2$. 
The former is true because we can prove by refutation that $\psi_1$ is unsatisfiable.
To do so, we simply resolve $P(a_{[0,1],1}) \in N_G$ with $(\neg P(a_{[0,1],1}) \vee Q(a_{[0,1],1})) \in N_G$ and $P(a_{[0,1],2}) \in N_G$ with $(\neg P(a_{[0,1],2}) \vee Q(a_{[0,1],2})) \in N_G$ to get $Q(a_{[0,1],1})$ and $Q(a_{[0,1],2})$.
Hence, $(Q(a_{[0,1],1}) \wedge \neg Q(a_{[0,1],2}))$ in $\phi'_1$ simplifies to false because of $Q(a_{[0,1],2})$, $(Q(a_{[0,1],2}) \wedge \neg Q(a_{[0,1],1}))$ in $\phi'_1$ simplifies to false because of $Q(a_{[0,1],1})$, and as a result $\phi'_1$ overall also simplifies to false.
Our second conjectures is not a consequence, i.e. $N \not\models \forall x, y.\phi_2$, because $\psi_2$ actually has a satisfying interpretation $\sigval$ with
$P^{\sigval} := \{a_{(-\infty,0),1}, a_{(-\infty,0),2}, a_{[0,1],1}, a_{[0,1],2}\}$ and
$Q^{\sigval} := \{a_{[0,1],1}, a_{[0,1],2}, a_{(1,2],1}\}$.
This assignment satisfies $\psi_2$ and constitutes a counter example for $N \models \forall x, y.\phi_2$ because it satisfies $N$ although it assigns $Q$ for one test point ($a_{(1,2],1}$) in $(1,2]$ to true and for the other ($a_{(1,2],2}$) to false.
Hence, $N$ is not uniform in the interval $(1,2]$.
\end{example}
\end{extendedonly}

\subsubsection{A Datalog Hammer for $\HBS(\SB)\PP$:}
The set $\mGnd_B(N)$ grows exponentially with regard to the maximum number of variables $n_C$ in any clause $(\Lambda \parallel C) \in N$, i.e. $O(|\mGnd_B(N)|) = O(|N|\cdot|B|^{n_C})$.
Since $B$ is large for realistic examples (e.g., in our examples the size of $B$ ranges from 15 to 1609 constants), the finite abstraction is often too large to be solvable in reasonable time. 
As an alternative approach, we propose a Datalog hammer for the Horn fragment of $\BS(\SB)\PP$ clause sets, called \HBS(\SB)\PP.
This hammer exploits the ideas behind the finite abstraction and will allow us to make the same ground deductions, but instead of grounding everything, we only need to (i)~ground the negated conjecture over our test points and (ii)~provide a set of ground facts that define which theory atoms are satisfied by our test points.
As a result, the hammered formula is much more concise and we need no actual theory reasoning to solve the formula.
In fact, we can solve the hammered formula by greedily resolving with all facts (from our set of clauses and returned as a result of this process) until this produces the empty clause---which would mean the conjecture is implied---or no more new facts---which would mean we have found a counter example.
(In practice, greedily applying resolution is not the best strategy and we recommend to use more advanced techniques for instance those used by a state-of-the-art Datalog reasoner.)

The Datalog hammer takes as input 
(i)~a \HBS(\SB)\PP{} clause set $N$ (where $S$ is the set of all positively grounded predicates in $N$) and 
(ii)~optionally a universal conjecture $\forall \bar{y}. P(\bar{y})$ where $P \not\in S$.
Restricting the conjecture to a single positive literal may seem like a drastic restriction, but we will later show that we can transform any universal conjecture into this form if it contains only positive atoms.
Given this input, the Datalog hammer first computes the same interval partition $\intvp$ and test point/constant set $B$ needed for the finite abstraction.
Then it computes an assignment $\beta$ for the constants in $B$ that corresponds to the interval partition, i.e. $a_{I,i}  \beta \in I$ and $a_{I,i}  \beta \neq a_{I,j}  \beta$ if $i \neq j$.
Next, it computes three clause sets that will make up the Datalog formula. 
The first set $\tren_N(N)$ is computed out of $N$ by replacing each theory atom $A$ in $N$ with a literal $P_A(\bar{x})$, where $\vars(A) = \vars(\bar{x})$ and $P_A$
is a fresh predicate.
This is necessary to eliminate all non-constant function symbols (e.g., $+,-$) in positively grounded theory atoms because Datalog does not support non-constant function symbols.
(It is possible to reduce the number of fresh predicates needed, e.g., by reusing the same predicate for two theory atoms that are equivalent up to variable renaming.)
The second set is empty if we have no universal conjecture or it contains the ground and negated version $\phi$ of our universal conjecture $\forall \bar{y}. P(\bar{y})$. Since we restricted the conjecture to a single positive literal, $\phi$ has the form $C_{\phi} \rightarrow \bot$, where $C_{\phi}$ contains all literals $P(\bar{y})  \rho$ for all groundings $\rho : \vars(\bar{y}) \rightarrow B$.
We cannot skip this grounding but the worst-case size of $C_\phi$ is $O(\mGnd_B(N)) = O(|B|^{n_{\phi}})$,
where $n_{\phi} = |\bar{y}|$, which is in our applications typically much smaller than the maximum number of variables $n_C$ contained in any clause in $N$.
The last set is denoted by $\tfacts(N,B)$ and contains a fact $\tren_N(A)$ for every ground theory atom $A$ contained in the theory part $\Lambda$ of a clause $(\Lambda \parallel C) \in \mGnd_B(N)$ such that $A \beta$ simplifies to true. (Alternatively, it is also possible to use a set of axioms and a smaller set of facts and let the Datalog reasoner compute all relevant theory facts for itself.)
The set $\tfacts(N,B)$ can be computed without computing $\mGnd_B(N)$ if we simply iterate over all theory atoms $A$ in all constraints $\Lambda$ of all clauses $(\Lambda \parallel C) \in N$ 
and compute all groundings $\tau: \vars(A) \rightarrow B$ such that $A \tau \beta$ simplifies to true. 
This can be done in time $O(\mu(n_v) \cdot n_L \cdot {|B|}^{n_v})$ and the resulting set $\tfacts(N,B)$ has worst-case size $O(n_A \cdot {|B|}^{n_v})$,
where $n_L$ is the number of literals in $N$, $n_v$ is the maximum number of variables $|\vars(A)|$ in any theory atom $A$ in $N$, $n_A$ is the number of different theory atoms in $N$, and $\mu(x)$ is the time needed to simplify a theory atom over $x$ variables to a variable bound.
Please note that already satifiability testing for $\BS$ clause is NEXPTIME-complete in general, and DEXPTIME-complete for the Horn case~\cite{Lewis80,Plaisted84}.
So when abstracting to a polynomially decidable clause set (ground $\HBS$) an exponential factor is unavoidable.

\begin{lemma}\label{lem:hammeringequivalence}
$N \wedge \exists \bar{y}. \neg P(\bar{y})$ is equisatisfiable to its hammered version $N_D = \tren_N(N) \cup \tfacts(N,B) \cup \{\phi\}$. $N$ is equisatisfiable to its hammered version $\tren_N(N) \cup \tfacts(N,B)$.
\end{lemma}
\begin{prooftoggle}
Let $\preds'$ be the set of new predicate symbols introduced by $\tren_N$. 
We will prove that $\psi = \mGnd_B(N) \cup \idef(B) \cup \{\phi\}$, the finite abstraction of $N \wedge \exists \bar{y}. \neg P(\bar{y})$, is equisatisfiable to $N_D = \tren_N(N) \cup \tfacts(N,B) \cup \{\phi\}$. Then we get from Lemma~\ref{lem:verificationequiv} that $N \wedge \exists \bar{y}. \neg P(\bar{y})$ is equisatisfiable to its hammered version. (The case for $N$ without conjecture works exactly the same.)\newline
$\Rightarrow$: Let $\sigval$ be an interpretation satisfying $\psi$. 
Then we can extend $\sigval$ over $\preds'$ so it also satisfies $N_D$. 
The extension sets exactly those arguments for $P_A \in \preds'$ to true that appear in $\tfacts(N,B)$, i.e. if $P_A \in \preds'$, then  $P_A^{\sigval} = \{\bar{a}^{\sigval} \mid P_A(\bar{a}) \in \tfacts(N,B)\}$.
As a result, $\sigval$ automatically satisfies $\tfacts(N,B)$ and $\sigval$ also trivially satisfies $\phi$ because it also appears in $\psi$.
Moreover, we can proof that $\sigval$ satisfies any clause $D \vee C \in \mGnd_B(\tren_N(N))$ by case distinction over the corresponding clause $\Lambda \parallel C \in \mGnd_B(N)$ with $\tren_N(\Lambda) = \neg D$: since $\sigval$ satisfies $\Lambda \parallel C$ (i)~either $\sigval$ satisfies $C$ or (ii)~$\sigval$ does not satisfy $\Lambda$ and therefore one of the atoms $P(\bar{a})$ in $\tren_N(\Lambda)$ does not appear in $\tfacts(N,B)$ by definition of $\tfacts$ and thus $D$ that contains $\neg P(\bar{a})$ is satisfied by $\sigval$.
Hence, $\sigval$ satisfies $\tren_N(N)$.\newline
$\Leftarrow$: 
Let $\sigval$ be an interpretation satisfying $N_D$. 
Let $\beta$ be the assignment for the constants in $B$ that was used in the construction of $N_D$ such that $a_{I,i} \beta \in I$ and $a_{I,i} \beta \neq a_{I,j} \beta$ if $i \neq j$.
Then there exists an interpretation $\sigval'$ that satisfies $\psi$.
$\sigval'$ interprets each constant $a_{I,i}$ in $B$ as $a_{I,i} \beta$ and 
each predicate $P \in \preds$ as $P^{\sigval'} = \{\bar{a} \beta \mid \bar{a} \in P^{\sigval}\}$.
By definition of $\beta$, $\sigval'$ satisfies $\idef(B)$.
As in the previous case, $\sigval'$ satisfies $\phi$ because $\phi$ appears in $N_D$ and $\sigval$ satisfies $N_D$.
Moreover, we can proof that $\sigval'$ satisfies any clause $(\Lambda \parallel C) \in \mGnd_B(N)$ by case distinction over the corresponding clause $(D \vee C) \in \mGnd_B(\tren_N(N))$ with $\tren_N(\Lambda) = \neg D$: since $\sigval$ satisfies $D \vee C$ (i)~either $\sigval$ satisfies $C$ and therefore $\sigval'$ satisfies $C$ or (ii)~$\sigval$ satisfies $D$ and therefore at least one of the atoms $P(\bar{a}) = \tren_N(A)$ with $A \in \Lambda$ does not appear in $\tfacts(N,B)$, which can only be that case if $A \beta$ simplifies to false and thus $\Lambda$ is not satisfied by $\sigval'$.
Hence, $\sigval$ also satisfies $\mGnd_B(N)$.
\end{prooftoggle}

Note that $\tren_N(N) \cup \tfacts(N,B) \cup \{\phi\}$ is actually a $\HBS$ clause set over a finite set of constants $B$ and not yet a Datalog input file.
It is well known that such a formula can be transformed easily into a Datalog problem by adding a nullary predicate Goal and adding it as a positive literal to any clause without a positive literal.
Querying for the Goal atom returns true if the $\HBS$ clause set was unsatisfiable and false otherwise.

\subsubsection{Positive Conjectures:}
One of the seemingly biggest restrictions of our Datalog hammer is that it only accepts universal conjectures over a single positive literal $\forall \bar{y}. P(\bar{y})$.
We made this restriction because it is the easiest way to guarantee that our negated and finitely abstracted goal takes the form of a Horn clause.
However, there is a way to express any positive universal conjecture --- i.e. any universal conjecture where all atoms have positive polarity --- as a universal conjecture over a single positive literal. (Note that any negative theory literal can be turned into a positive theory literal by changing the predicate symbol, e.g., $\neg (x \leq 5) \equiv (x > 5)$.)
Similarly as in a typical first-order CNF transformation~\cite{NonnengartW01}, we can simply rename all subformulas, i.e. recursively replace all subformulas with some some fresh predicate symbols 
and add suitable Horn clause definitions for these new predicates to our clause set $N$.
\begin{notinextended}
A detailed algorithm for this flattening process and a proof of equisatisfiability can be found in the extended version of this paper.
Using the same technique, we can also express any positive existential conjecture --- i.e. any existential conjecture where all atoms have positive polarity --- as additional clauses in our set of input clauses $N$.
\end{notinextended}

\begin{extendedonly}
Let $\forall \bar{y}. \phi'$ be a universal conjecture where all atoms have positive polarity. 
Then we define the functions $\rflat(\phi')$ and $\pflat(\phi')$ recursively as follows:
$\pflat(\phi')$ returns an atom $P_{\phi'}(\bar{x})$ over a fresh predicate $P_{\phi'}$ for any formula $\phi'$ with $\vars(\phi')=\vars(\bar{x})$ that is not just a free first-order atom and otherwise the atom itself.
$\rflat(\phi')$ on the other hand introduces a set of new rules that define the fresh predicates $P_{\phi'}$: $\rflat(\phi') := \{(\pflat(\phi'_1), \ldots, \pflat(\phi'_m) \rightarrow \pflat(\phi'))\} \cup \rflat(\phi'_1) \cup \ldots \cup \rflat(\phi'_m)$ if $\phi' = \phi'_1 \wedge \ldots \wedge \phi'_m$, $\rflat(\phi') := \{(\pflat(\phi'_1) \rightarrow \pflat(\phi')), \ldots, (\pflat(\phi'_m) \rightarrow \pflat(\phi'))\}  \cup \rflat(\phi'_1) \cup \ldots \cup \rflat(\phi'_m)$ if $\phi' = \phi'_1 \vee \ldots \vee \phi'_m$, 
$\rflat(\phi') := \{(\Lambda \parallel \rightarrow \pflat(\phi'))\}$ if $\phi' = \Lambda$ is a conjunction of theory atoms, and 
$\rflat(\phi') := \emptyset$  if $\phi'$ is a free first-order atom.

\begin{lemma}\label{lem:positiveconjectureflattening}
Let $\sigval$ be an interpretation that satisfies $N$.
Let $\tau : \vars(\phi^*) \rightarrow \Real$ be a grounding for $\phi'$.
Then $\sigval \models \phi' \tau$ if and only if $(\sigval \wedge \rflat(\phi')) \models \pflat(\phi') \tau$.
\end{lemma}
\begin{prooftoggle}
$\Rightarrow$: Assume $\sigval$ is an interpretation that satisfies $\sigval \models \phi' \tau$.
Then we show by induction that $\sigval \wedge \rflat(\phi') \models \pflat(\phi^*) \tau$ for all subformulas $\phi^*$ of $\phi'$, where $\sigval \models \phi^* \tau$.
Case 1: if $\phi^*$ is a free first-order atom, then $(\sigval \wedge \rflat(\phi')) \models \pflat(\phi^*) \tau$ because $\pflat(\phi^*) = \phi^* \tau$.
Case 2: if $\phi^* = \Lambda$ is a conjunction of theory atoms, then $(\Lambda \parallel \rightarrow \pflat(\phi^*))\in \rflat(\phi')$ and $\sigval \models \phi^* \tau$ entails $(\sigval \wedge \rflat(\phi')) \models \pflat(\phi^*) \tau$.
Case 3: if $\phi^* = \phi^*_1 \vee \ldots \vee \phi^*_m$, then there must exist a $\phi^*_j$ with $\sigval \models \phi^*_j \tau$ and by induction hypothesis $(\sigval \wedge \rflat(\phi')) \models \pflat(\phi^*_j) \tau$. Together with 
$(\pflat(\phi^*_j) \rightarrow \pflat(\phi^*))\in \rflat(\phi')$ this means $(\sigval \wedge \rflat(\phi')) \models \pflat(\phi^*) \tau$.
Case 4: if $\phi^* = \phi^*_1 \wedge \ldots \wedge \phi^*_m$, then $\sigval \models \phi^*_j \tau$ for all $\phi^*_j$ and by induction hypothesis $(\sigval \wedge \rflat(\phi')) \models \pflat(\phi^*_j) \tau$ for all $\phi^*_j$. Together with 
$(\pflat(\phi^*_1), \ldots, \pflat(\phi^*_m) \rightarrow \pflat(\phi^*))\in \rflat(\phi')$ this means $(\sigval \wedge \rflat(\phi')) \models \pflat(\phi^*) \tau$.\newline
$\Leftarrow$: Assume $\sigval$ is an interpretation that satisfies $(\sigval \wedge \rflat(\phi')) \models \rflat(\phi')\tau$. 
Then we show by induction that $\sigval \models \phi^* \tau$ for all subformulas $\phi^*$ of $\phi'$, where $(\sigval \wedge \rflat(\phi')) \models \pflat(\phi^*) \tau$.
Case 1: if $\phi^*$ is a free first-order atom, then $\pflat(\phi^*) = \phi^*$ only appears as a negative literal in $\rflat(\phi')$. Hence, $\sigval \models \phi^* \tau$.
Case 2: if $\phi^* = \Lambda$ is a conjunction of theory atoms, then $(\Lambda \parallel \rightarrow \pflat(\phi^*))\in \rflat(\phi')$ is the only clause in $N \cup \rflat(\phi')$, where $\pflat(\phi^*)$ appears positively.
Hence, $\sigval \models \phi^* \tau$.
Case 3: if $\phi^* = \phi^*_1 \vee \ldots \vee \phi^*_m$, then there only exist $m$ clauses in $\rflat(\phi')$ where $\pflat(\phi')$ is positive. This means $(\sigval \wedge \rflat(\phi')) \models \pflat(\phi^*) \tau$ can only be true if there exist a rule $(\pflat(\phi^*_j) \rightarrow \pflat(\phi^*))\in \rflat(\phi')$ with $(\sigval \wedge \rflat(\phi')) \models \pflat(\phi^*_j) \tau$ and by induction hypothesis $\sigval \models \phi^*_j \tau$. However, this also means $\sigval \models \phi^* \tau$.
Case 4: if $\phi^* = \phi^*_1 \wedge \ldots \wedge \phi^*_m$, then $(\pflat(\phi^*_1), \ldots, \pflat(\phi^*_m) \rightarrow \pflat(\phi^*))\in \rflat(\phi')$ is the only clause in $\rflat(\phi')$ where $\pflat(\phi')$ is positive. This means $(\sigval \wedge \rflat(\phi')) \models \pflat(\phi^*) \tau$ can only be true if $(\sigval \wedge \rflat(\phi')) \models \pflat(\phi^*_j) \tau$ for all $\phi^*_j$ and by induction hypothesis $\sigval \models \phi^*_j \tau$. However, this also means $\sigval \models \phi^* \tau$.
\end{prooftoggle}

\begin{corollary}\label{cor:positiveunivconjectureflattening}
$N \models (\forall \bar{y}. \phi')$ is equivalent to $(N \cup \rflat(\phi')) \models (\forall \bar{y}. \pflat(\phi'))$.
\end{corollary}

Using the same technique, we can also express any positive existential conjecture --- i.e. any existential conjecture where all atoms have positive polarity --- as additional clauses in our set of input clauses $N$.

\begin{corollary}\label{cor:positiveexistconjectureflattening}
$N \models (\exists \bar{y}. \phi')$ is equivalent to $N \cup \rflat(\phi') \cup (\pflat(\phi') \rightarrow \bot)$ is unsatisfiable.
\end{corollary}
\end{extendedonly}

\begin{extendedonly}
As with the first hammer, we also finish the presentation of our second hammer by applying it to some examples. 

\begin{example}
For our examples, we choose the same set of clauses as in the previous example\newline
\centerline{$N := \{0 \leq x, x \leq 2 \parallel \neg P(x) \vee Q(x), \quad x \leq 1 \parallel P(x), \quad x > 1 \parallel \neg P(x)\}$}
Although this set of clauses belongs to the horn fragment, we cannot handle the previous two conjectures with our Datalog hammer because they are not positive (both contain the negative literal $\neg Q(x)$).
Instead, we will check two different conjectures
$\phi_3 := 0 \leq x, x \leq 1 \parallel Q(x)$ and $\phi_4 := 0 \leq x, x \leq 2 \parallel Q(x)$. 
$N \models \forall x. \phi_3$ is true if every interpretation $\sigval$ satisfying $N$ interprets $Q(a)$ as true for all $a \in [0,1]$.
$N \models \forall x. \phi_4$ is true if every interpretation $\sigval$ satisfying $N$ interprets $Q(a)$ as true for all $a \in [0,2]$.
Before we can apply the Datalog hammer, we have to flatten our positive conjectures to unit clauses.
Note that $\phi_3$ written via boolean operators instead of our $\parallel$-notation looks as follows: $0 > x \vee x > 1 \vee Q(x)$.
Hence, $\pflat(\phi_3) := R(x)$ and $\rflat(\phi_3) = \{0 > x \parallel R(x), \; x > 1 \parallel R(x), \; Q(x) \rightarrow R(x)\}$.
Similarly, $\pflat(\phi_4) := S(x)$ and $\rflat(\phi_4) = \{0 > x \parallel S(x), \; x > 2 \parallel S(x), \; Q(x) \rightarrow S(x)\}$.
Next, we combine these new clause sets with our initial clause set $N$ 
to get $N_3 := N \cup \rflat(\phi_3)$ and $N_4 := N \cup \rflat(\phi_4)$.
By Corollary~\ref{cor:positiveunivconjectureflattening}, this means that $N \models \forall x. \phi_3$ and $N \models \forall x. \phi_4$ are equivalent to $N_3 \models \forall x. R(x)$ and $N_4 \models \forall x. S(x)$, respectively.
The latter two are now also in a format to which we can apply the Datalog hammer.

As the first step of our Datalog hammer, 
we have to compute the interval partition for both examples.
Note, however, that the variable bounds in $\rflat(\phi_3)$ and $\rflat(\phi_4)$ appear in negated form in $N$, e.g. $0 > x$ appears in $\rflat(\phi_3)$ and $0 \leq x \equiv \neg (0 > x)$ appears in $N$.
This means $N_3$ and $N_4$ result in the same interval endpoints as $N$ alone.
The latter we have already determined in the previous example as $\intvp = \{(-\infty,0),[0,1],(1,2],(2,\infty)\}$.
In contrast to our previous example, our conjectures contain only one variable, 
so we need only one test point/constant for each interval.
Therefore, $B = \{a_{(-\infty,0),1}, a_{[0,1],1}, a_{(1,2],1}, a_{(2,\infty),1}\}$.
As our assignment $\beta$ for the constants, we can simply pick random points from the respective intervals, e.g. $a_{(-\infty,0),1} \beta = -1$, $a_{[0,1],1} \beta = 0$, $a_{(1,2],1} \beta = 2$, and $a_{(2,\infty),1} \beta = 3$.

In the next step of the Datalog hammer, we have to replace the inequalities in our clause sets by fresh predicates.
The inequalities in our examples are 
$0 \leq x$, $x \leq 2$, $x \leq 1$, $x > 1$, $0 > x$, and $x > 2$; 
we choose as their respective fresh predicates $T_{0 \leq x}(x)$, $T_{x \leq 2}(x)$, $T_{x \leq 1}(x)$, $T_{x > 1}(x)$, $T_{0 > x}(x)$, and $T_{x > 2}(x)$.
As a result,
our sets of clauses after renaming look as follows\newline
\centerline{$\begin{array}{l l}
\tren_{N_3}(N_3) := \{
&\neg T_{0 \leq x}(x) \vee T_{x \leq 2}(x) \vee \neg P(x) \vee Q(x), \quad
\neg T_{x \leq 1}(x) \vee P(x), \\
&\neg T_{x > 1}(x) \vee \neg P(x), \; \neg T_{0 > x}(x) \vee R(x), \quad \neg T_{x > 1}(x) \vee R(x), \\
&\neg Q(x) \vee R(x)\}
\end{array}$}
\centerline{$\begin{array}{l l}
\tren_{N_4}(N_4) := \{
&\neg T_{0 \leq x}(x) \vee T_{x \leq 2}(x) \vee \neg P(x) \vee Q(x), \quad 
\neg T_{x \leq 1}(x) \vee P(x), \\
&\neg T_{x > 1}(x) \vee \neg P(x), \quad
\neg T_{0 > x}(x) \vee S(x), \; \neg T_{x > 2}(x) \vee S(x), \\
&\neg Q(x) \vee S(x)\}
\end{array}$}
The groundings of our conjectures are\newline
\centerline{$\phi'_3 := \neg R(a_{(-\infty,0),1}) \vee \neg R(a_{[0,1],1}) \vee \neg R(a_{(1,2],1}) \vee \neg R(a_{(2,\infty),1})$}
\centerline{$\phi'_4 := \neg S(a_{(-\infty,0),1}) \vee \neg S(a_{[0,1],1}) \vee \neg S(a_{(1,2],1}) \vee \neg S(a_{(2,\infty),1})$} 
so simply the disjunction of all negative instances of our conjecture predicate over our set of test points $B$.
The sets of theory facts are \newline
\centerline{$\begin{array}{l l}
\tfacts(N_3,B) := \{
&T_{0 \leq x}(a_{[0,1],1}), T_{0 \leq x}(a_{(1,2],1}), T_{0 \leq x}(a_{(2,\infty),1}), 
T_{x \leq 2}(a_{(-\infty,0),1}),\\ &T_{x \leq 2}(a_{[0,1],1}), T_{x \leq 2}(a_{(1,2],1}),
T_{x \leq 1}(a_{(-\infty,0),1}), T_{x \leq 1}(a_{[0,1],1}),\\
&T_{x > 1}(a_{(1,2],1}), T_{x > 1}(a_{(2,\infty),1}),
T_{0 > x}(a_{(-\infty,0),1})
\}
\end{array}
$}
\centerline{$\begin{array}{l l}
\tfacts(N_4,B) := \{
&T_{0 \leq x}(a_{[0,1],1}), T_{0 \leq x}(a_{(1,2],1}), T_{0 \leq x}(a_{(2,\infty),1}),
T_{x \leq 2}(a_{(-\infty,0),1}),\\ &T_{x \leq 2}(a_{[0,1],1}), T_{x \leq 2}(a_{(1,2],1}),
T_{x \leq 1}(a_{(-\infty,0),1}), T_{x \leq 1}(a_{[0,1],1}),\\
&T_{x > 1}(a_{(1,2],1}), T_{x > 1}(a_{(2,\infty),1}),
T_{0 > x}(a_{(-\infty,0),1}),
T_{x > 2}(a_{(2,\infty),1})
\}
\end{array}$}
The last step of our Datalog hammer is to combine these sets of clauses 
into one big set of horn clauses for each conjecture:
$N_D^3 := \tren_{N_3}(N_3) \cup \{\phi'_3\} \cup \tfacts(N_3,B)$ and $N_D^4 := \tren_{N_4}(N_4) \cup \{\phi'_4\} \cup \tfacts(N_4,B)$.

We end this example by proving that 
(i)~the conjecture $\phi_3$ is actually a consequence of $N$ and
(ii)~the conjecture $\phi_4$ is not a consequence of $N$.
We prove that the conjecture $\phi_3$ is actually a consequence of $N$ by deriving the empty clause from $N_D^3$, which proves that $N_D^3$ is unsatisfiable (and that there exists no counter example).
As our first resolution step, we will simply resolve with all $\tfacts$.
The result will be similar to what we got, when we simplified the ground clauses in our first hammer with the help of the inequalities in $\idef(B)$ and theory reasoning; 
we will get rid of all the renamed theory literals:\newline
\centerline{$
\begin{array}{l l}
N_G^3 := \{
&\neg P(a_{[0,1],1}) \vee Q(a_{[0,1],1}), \quad
\neg P(a_{(1,2],1}) \vee Q(a_{(1,2],1}), \\
&P(a_{(-\infty,0),1}),  \quad
P(a_{[0,1],1}),  \quad
\neg P(a_{(1,2],1}),  \quad
\neg P(a_{(2,\infty),1}),  \\
&R(a_{(-\infty,0),1}),  \quad
R(a_{(1,2],1}),  \quad
R(a_{(2,\infty),1})
\}
\end{array}
$}
Next, we resolve $P(a_{[0,1],1})$ with $\neg P(a_{[0,1],1}) \vee Q(a_{[0,1],1})$ to get $Q(a_{[0,1],1})$, followed by $Q(a_{[0,1],1})$ with $Q(x) \vee R(x)$ to get $R(a_{[0,1],1})$.
The final step is to resolve $\phi'_3$ with $R(a_{[0,1],1})$ and $R(a_{(-\infty,0),1}), R(a_{(1,2],1}), R(a_{(2,\infty),1}) \in N_G^3$.
The result is the empty clause.

Our second conjecture $\phi_4$ is not a consequence, i.e. $N \not\models \forall x.\phi_4$, because $N_D^3$ actually has a satisfying interpretation $\sigval$ with
$P^{\sigval} := \{a_{(-\infty,0),1}, a_{[0,1],1}\}$,
$Q^{\sigval} := \{a_{[0,1],1}\}$ and
$R^{\sigval} := \{a_{(-\infty,0),1}, a_{[0,1],1}, a_{(2,\infty),1}\}$.
\end{example}
\end{extendedonly}


\section{Two Supervisor Case Studies} \label{sec:super}

We consider two supervisor case studies: a lane change assistant and
the ECU of a supercharged combustion engine; both using the architecture in \figref{supervisor-arch}.


\myparagraph{Lane Assistant:}
\label{subsec:lane} 
This use case focuses on the lane changing maneuver in autonomous driving scenario \ie the safe \emph{lane} selection and the \emph{speed}.
We run two variants of software processing units (updated and certified) in parallel with a supervisor.
The variants are connected to different sensors that capture the state of the freeway such as video or LIDAR signal sensors.
The variants process the sensors' data and suggest the safe lanes to change to in addition to the evidence that justify the given selection.
The supervisor is responsible for the selection of which variant output to forward to other system components \ie the execution units (actuators) that perform the maneuver.
Variants categorize the set of available actions for each time frame into \emph{safe/unsafe} actions and provide \emph{explications}.
The supervisor collects the variants output and processes them to reason about  
\begin{inparaenum} [(a)]
\item if enough evidence is provided by the variants to consider actions safe
\item find the actions that are considered safe by all variants. 
\end{inparaenum}

Variants formulate their explications as \emph{facts} using first-order predicates. 
The supervisor uses a set of logical \emph{rules} formulated in $\BS(\SB)\PP$ to reason about the suggestions and the explications (see \lstref{lc-facts-rules}).
In general, the rules do not belong to the $\BS(\SB)\PP$ fragment, e.g., the atom $=(xh1,-(xes,1))$ includes even an arithmetic
calculation. However, after grounding with the facts of the formalization, only simple bounds remain.

\begin{lstlisting}[escapeinside={(*}{*)}, mathescape=true, caption=The rules snippets for the lane changing use case in $\BS(\SB)\PP$., label={lst:lc-facts-rules}]
(*\speccomment{\#\# Exclude actions per variant if safety disproved or declared unsafe.}*)
SuggestionDisproven(xv, xa), VariantName(xv) -> ExcludedAction(xv, xa). (*\label{line:s-rules}*) (*\label{line:safety-disproven}*)
VariantName(xv), LaneNotSafe(xv, xl, xa)     -> ExcludedAction(xv, xa). (*\label{line:not-safe-declared}*)
(*\speccomment{\#\#  Exclude actions for all variants if declared unsafe by the certified}*)
CertifiedVariant(xv1), UpdatedVariant(xv2), LaneNotSafe(xv1, xl, xa) (*\label{line:s-certified-excluded-across-all-variants}*)
  -> ExcludedAction(xv2,xa).(*\label{line:e-certified-excluded-across-all-variants}*)

(*\speccomment{\#\# A safe action is disproven}*)
SafeBehindDisproven(xv, xenl, xecl, xecs, xes, xa), LaneSafe(xv, xl, xa),
  SuggestedAction(xv, xa)  -> SuggestionDisproven(xv, xa).
SafeFrontDisproven(xv, xenl, xecl, xecs, xes, xa),  LaneSafe(xv, xl, xa), 
  SuggestedAction(xv, xa)  -> SuggestionDisproven(xv, xa).

(*\speccomment{\#\# Unsafe left lane: speed decelerated and unsafe distance front}*)
>(xh1, xfd), !=(xecl, xenl),  =(xh1,-(xes,1)) || (*\label{line:s-safe-front-disproven}*) 
  LaneSafe(xv, xenl, adecelerateleft), EgoCar(xv, xecl, xecs, xes), 
  DistanceFront(xv, xenl, xofp, xfd, adecelerateleft), 
  SpeedFront(xv, xenl, xofp, xofs, adecelerateleft) 
  -> SafeFrontDisproven(xv, xenl, xecl, xecs, xes, adecelerateleft).  (*\label{line:e-safe-front-disproven}*) (*\label{line:e-rules}*)
\end{lstlisting}

\mysubparagraph{Variants explications:}
The \code{SuggestedAction} predicate encodes the actions suggested by the variants. 
\code{LaneSafe} and \code{LaneNotSafe} specify the lanes that are safe/unsafe to be used with the different actions.
\code{DistanceFront} and \code{DistanceBehind} provide the explications related to the obstacle position, while their speeds are \code{SpeedFront} and \code{SpeedBehind}.
\code{EgoCar} predicate reports the speed and the position of the ego vehicle.

\mysubparagraph{Supervisor reasoning:}
To select a safe action, the supervisor must exclude all unsafe actions.
The supervisor considers actions to be excluded per variant (\code{Ex\-cludedAction}) if
\begin{inparaenum} [(a)]
\item \code{SuggestionDisproven}; the variant fails to prove that the suggested action is safe (\linref{safety-disproven}), or
\item the action is declared unsafe (\linref{not-safe-declared}). 
\end{inparaenum}
The supervisor declares an action to be excluded cross all variants if the certified variant declares it unsafe (\linreftwo{s-certified-excluded-across-all-variants}{e-certified-excluded-across-all-variants}).
To consider an action as \code{SuggestionDisproven}, the supervisor must check for each \code{LaneSafe} the existence of unsafe distances between the ego vehicle in the given lane and the other vehicles approaching either from behind (\code{SafeBehindDisproven}) or in front (\code{SafeFrontDisproven}).
The rule \code{SafeFrontDisproven} (\linreftwo{s-safe-front-disproven}{e-safe-front-disproven}) checks in the left lane, if using the ego vehicle decelerated speed (\code{=(xh1,-(xes,1))}) the distance between the vehicles is not enough (\code{>(xh1, xfd)}).
The supervisor checks \code{ExcludeAction} for all variants.
If all actions are excluded, the supervisor uses an emergency action as no safe action exists.
Otherwise, selects a safe action from the not-excluded actions suggested by the updated variant, if not found, by the certified.


\myparagraph{ECU:}
\label{subsec:ecu}
The GM LSJ Ecotec engine (\url{https://en.wikipedia.org/wiki/GM_Ecotec_engine}) is a supercharged
combustion engine that was almost exclusively deployed in the US, still some of those run also
in Europe. The main sensor inputs of the LSJ ECU 
consist of an inlet air pressure and temperature sensor (in KPa and in degree Celsius), a speed sensor (in Rpm),
a throttle pedal sensor, a throttle sensor, a coolant temperature sensor, oxygen sensors, a knock sensor, and its main
actuators controlling the engine are ignition and injection timing, and throttle position. For the experiments
conducted in this paper we have taken the routines of the LSJ ECU that compute ignition and injection timings
out of inlet air pressure, inlet air temperature, and engine speed. For this part of the ECU this is a
two stage process where firstly, basic ignition and injection timings are computed out of engine speed and
inlet air pressure and secondly, those are adjusted with respect to inlet air temperature. The properties
we prove are safety properties, e.g., certain injection timings are never generated and also invariants, e.g.,
the ECU computes actuator values for all possible input sensor data and they are unique.
Clause~\ref{eq_bs_lra_ecu_nonground}, page~\pageref{eq_bs_lra_ecu_nonground}, is an actual clause
from the ECU case study computing the base ignition timing.
\begin{extendedonly}
  The adjustments are then done with respect to inlet temperature values. Basically, if the inlet
  temperature exceeds a certain value, typically around $80^\circ$C, then the pre-ignition is reduced
  in order to prevent a too early burning of the fuel that might damage the engine.
\begin{center}\begin{tabular}{l}
                $z_1\geq y_1, x_1\leq x, x < x_2 || \neg\text{IgnDeg1}(xtp, xtr, xp, xrpm, z_1) \lor \neg\text{Tmp}(x,x)\quad\lor$ \\
                $\qquad \neg\text{TempIgnTabl}e(x_1,x_2,y_1,y_2) \lor \text{IndnDeg2}(xtp, xtr, x, xp, xrpm, x, y_ 2)$
              \end{tabular}\end{center}
            A final third iteration then considers the knock sensor. If knock is detected consistently over a certain period of time,
            then the pre-ignition is reduced even further. This iteration is not yet contained in our formalization.
\end{extendedonly}



\section{Implementation and Experiments} \label{sec:experiments}

We have implemented the Datalog hammer into our $\BS(\LRA)$ system SPASS-SPL and combined it with the Datalog reasoner Rulewerk.
The resulting toolchain is the first implementation of a decision procedure for \HBS(\SB) with positive conjectures.

\myparagraph{SPASS-SPL} 
is a new system for $\BS(\LRA)$ based on some core libraries of the first-order
theorem prover SPASS~\cite{WeidenbachEtAlSpass2009} and including the CDCL(LA) solver SPASS-SATT~\cite{BrombergerEtAl19} for mixed linear arithmetic.
Eventually, SPASS-SPL will include a family of reasoning techniques for $\BS(\LRA)$ including SCL(T)~\cite{BrombergerFW21},
hierarchic superposition~\cite{BachmairGanzingerEtAl94,BaumgartnerWaldmann19} and hammers to various logics.
Currently, it comprises the Datalog hammer described in this paper and hierarchic UR-resolution~\cite{McCharenOverbeekEtAl76} (Unit Resulting resolution)
which is complete for pure $\HBS(\LRA)$.
The Datalog hammer can produce the clause format used in the Datalog system \emph{Rulewerk} (described below), but also the SPASS first-order logic clause format that can then
be translated into the first-order TPTP library~\cite{Sutcliffe17} clause format.
Moreover, it can be used as a translator from our own input language into the SMT-LIB 2.6 language~\cite{BarFT-RR-17} and the CHC competition format~\cite{Ruemmer20CHC}.

Note that our implementation of the Datalog hammer is of prototypical nature. 
It cannot handle positively grounded theory atoms beyond simple bounds, unless they are variable comparisons (i.e., $x \LAOP y$ with $\LAOP \in \{\leq, <, \neq, =, >, \geq\}$).
Moreover, positive universal conjectures have to be flattened until they have the form $\Lambda \parallel P(\bar{x})$.
On the other hand, we already added some improvements,
e.g., we break/eliminate symmetries in the hammered conjecture and we exploit the theory atoms $\Lambda$ in a universal conjecture $\Lambda \parallel P(\bar{x})$ so the hammered conjecture contains only groundings for $P(\bar{x})$ that satisfy $\Lambda$.

\myparagraph{Rulewerk}
(formerly \emph{VLog4j}) is a rule reasoning toolkit that consists of a Java API and an
interactive shell \cite{Rulewerk2019}. Its current main reasoning back-end is the rule engine \emph{VLog} \cite{UJK:VLog2016}, which supports
Datalog and its extensions with stratified negation and existential quantifiers, respectively.
VLog is an in-memory reasoner that is optimized for efficient use of resources, and has been
shown to deliver highly competitive performance in benchmarks \cite{UKJDC18:VLogSystemDescription}.

We have not specifically optimized VLog or Rulewerk for this work, but we have tried to select Datalog
encodings that exploit the capabilities of these tools. The most notable impact was observed for the
encoding of universal conjectures. A direct encoding of (grounded) universal claims in Datalog
leads to rules with many (hundreds of thousands in our experiments) ground atoms as their precondition. Datalog reasoners (not just
VLog) are not optimized for such large rules, but for large numbers of facts.
An alternative encoding in plain Datalog would therefore specify the expected atoms as facts and
use some mechanism to iterate over all of them to check for goal. To accomplish this iteration, the facts
that require checking can be endowed with an additional identifier (given as a parameter), and an auxiliary
binary successor relation can be used to specify the iteration order over the facts. This approach requires only
few rules, but the number of rule applications is proportional to the number of expected facts.

In Rulewerk/VLog, we can encode this in a simpler way using negation.
Universal conjectures require us to evaluate ground queries of
the form $\textit{entailed}(\bar{c}_1)\wedge\ldots\wedge\textit{entailed}(\bar{c}_\ell)$, 
where each $\textit{entailed}(\bar{c}_i)$ represents one grounding of our conjecture over our set of test points.
If we add facts $\textit{expected}(\bar{c}_i)$ for the constant vectors $\bar{c}_1,\ldots,\bar{c}_\ell$,
we can equivalently use a smaller (first-order) query 
$\forall\bar{x}.(\textit{expected}(\bar{x})\to\textit{entailed}(\bar{x}))$, which 
in turn can be written as
$\neg\big(\exists\bar{x}.(\textit{expected}(\bar{x})\wedge\neg\textit{entailed}(\bar{x}))\big)$.
This can be expressed in Datalog with negation and the rules $\textit{expected}(\bar{x})\wedge\neg\textit{entailed}(\bar{x}) \to \textit{missing}$ and $\neg\textit{missing} \to \textit{Goal}$, where \textit{Goal} encodes that the query matches.
This use of negation is \emph{stratified}, i.e., not entwined with recursion \cite{Alice}.
Note that stratified negation is a form of non-monotonic negation, so we can no longer read such rules as first-order
formulae over which we compute entailments. Nevertheless, implementation is simple and stratified negation is a widely
supported feature in Datalog engines, including Rulewerk.
The encoding is particularly efficient since the rules using negation are evaluated only once.

\begin{figure}[t]
\centering
  {\scriptsize
  \setlength{\tabcolsep}{1pt}
    \begin{tabular}{|l|c|c|r|r|r|r||r|r|r|r||r|r|r|r|}
    \hline
        \multicolumn{1}{|c|}{Problem} & Q & Status & \multicolumn{1}{|c|}{$X$} & \multicolumn{1}{|c|}{$Y$} & \multicolumn{1}{|c|}{$B$} & \multicolumn{1}{|c||}{Size} & \multicolumn{1}{|c|}{t-time} & \multicolumn{1}{|c|}{h-time} & \multicolumn{1}{|c|}{p-time} & \multicolumn{1}{|c||}{r-time} & \multicolumn{1}{|c|}{vampire} & \multicolumn{1}{|c|}{spacer} & \multicolumn{1}{|c|}{z3} & \multicolumn{1}{|c|}{cvc4} \\ \hline\hline
        lc\_e1 & $\exists$ & true & 9 & 3 & 19 & 12/30 & 0.2 & 0.0 & 0.1 & 0.1 &       0.0        &0.0 & 0.0 & 0.0 \\ \hline
        lc\_e2 & $\exists$ & false & 9 & 3 & 17 & 13/27 & 0.2 & 0.0 & 0.1 & 0.1 &      0.0        &0.1 & timeout & timeout \\ \hline
        lc\_e3 & $\exists$ & false & 9 & 3 & 15 & 12/22 & 0.2 & 0.0 & 0.1 & 0.1 &      0.0        &0.0 & timeout & timeout \\ \hline
        lc\_e4 & $\exists$ & true & 9 & 3 & 21 & 12/35 & 0.2 & 0.0 & 0.1 & 0.1 &       0.0        &0.0 & 0.0 & 0.1 \\ \hline
        lc\_u1 & $\forall$ & false & 9 & 2 & 29 & 12/25 & 0.2 & 0.0 & 0.1 & 0.1 &      0.0        & N/A & timeout & timeout \\ \hline
        lc\_u2 & $\forall$ & false & 9 & 2 & 26 & 12/25 & 0.2 & 0.0 & 0.1 & 0.1 &      0.0        & N/A & timeout & timeout \\ \hline
        lc\_u3 & $\forall$ & true & 9 & 2 & 23 & 12/22 & 0.2 & 0.0 & 0.1 & 0.1 &       0.0        &N/A & 0.0 & 0.1 \\ \hline
        lc\_u4 & $\forall$ & false & 9 & 2 & 32 & 12/33 & 0.2 & 0.0 & 0.1 & 0.1 &      0.0        &N/A & timeout & timeout \\ \hline
        ecu\_e1 & $\exists$ & false & 10 & 6 & 311 & 27/649 & 1.1 & 0.1 & 0.3 &    0.7 &  0.5    &0.1 & timeout & timeout \\ \hline
        ecu\_e2 & $\exists$ & true & 10 & 6 & 311 & 27/649 & 1.1 & 0.1 & 0.3 &     0.7 &  0.5    &0.1 & 2.4 & 0.4 \\ \hline
        ecu\_u1 & $\forall$ & true & 11 & 1 & 310 & 27/651 & 1.1 & 0.1 & 0.3 &     0.7 &  94.6   & N/A & 145.2 & 0.3 \\ \hline
        ecu\_u2 & $\forall$ & false & 11 & 1 & 310 & 27/651 & 1.1 & 0.1 & 0.3 &    0.7 &  80.7   & N/A & timeout & timeout \\ \hline
        ecu\_u3 & $\forall$ & true & 9 & 2 & 433 & 27/1291 & 1.0 & 0.1 & 0.5 &      0.4 & 12.0    & N/A & 209.7 & 0.1 \\ \hline
        ecu\_u4 & $\forall$ & true & 9 & 2 & 1609 & 26/20459 & 12.4 & 2.9 &  3.2 & 6.3 & 526.5   & N/A & 167.7 & 0.1 \\ \hline
        ecu\_u5 & $\forall$ & true & 10 & 3 & 629 & 28/17789 & 22.6 & 0.7 &  2.1 & 19.8 & timeout     & N/A & timeout & timeout \\ \hline
        ecu\_u6 & $\forall$ & false & 10 & 3 & 618 & 27/15667 & 11.6 & 0.7 & 1.7 & 9.1 &  timeout     & N/A & timeout & timeout \\ \hline
    \end{tabular}
  }
  \caption{Benchmark results and statistics}
  \label{fig:toolchainresults}
\end{figure}

\myparagraph{Benchmark Experiments}
To test the efficiency of our toolchain, 
we ran benchmark experiments on the two real world $\HBS(\SB)\PP$ supervisor verification conditions.
The two supervisor use cases are described in Section~\ref{sec:super}.
The names of the problems are formatted so the lane change assistant examples start with lc and the ECU examples start with ecu.
The lc problems with existential conjectures test whether an action suggested by an updated variant is contradicted by a certified variant.
The lc problems with universal conjectures test whether an emergency action has to be taken because we have to exclude all actions for all variants.
The ecu problems with existential conjectures test safety properties, e.g., whether a computed actuator value is never outside of the allowed safety bounds.
The ecu problems with universal conjectures test whether the ecu computes an actuator value for all possible input sensor data.
Our benchmarks are prototypical for the complexity of $\HBS(\SB)$ reasoning in that they cover all abstract relationships between
conjectures and $\HBS(\SB)$ clause sets. With respect to our two case studies we have many more examples showing respective characteristics.
We would have liked to run benchmarks from other sources too, 
but we could not find any suitable $\HBS(\SB)$ problems in the SMT-LIB or CHC-COMP benchmarks.

For comparison, we also tested several state-of-the-art theorem provers for related logics (with the best settings we found): the satisfiability modulo theories (SMT) solver \emph{cvc4-1.8}~\cite{BarrettCDHJKRT:11} with settings \verb|--multi-trigger-cache| \verb|--full-saturate-quant|; the SMT solver \emph{z3-4.8.10}~\cite{deMouraBjorner:08} with its default settings;
the constrained horn clause (CHC) solver \emph{spacer}~\cite{KomuravelliGC14} with its default settings;
and the first-order theorem prover \emph{vampire-4.5.1}~\cite{RiazanovVoronkov02} with settings \verb|--memory_limit 8000| \verb|-p off|, i.e., with memory extended to 8GB and without proof output.

For the experiments, we used a Debian Linux server with 32 Intel Xeon Gold 6144 (3.5 GHz) processors and 754 GB RAM. 
Our toolchain employs no parallel computing, except for the java garbage collection.
The other tested theorem provers employ no parallel computing at all.
Each tool got a time limit of 40 minutes for each problem.

The table in Fig.~\ref{fig:toolchainresults} lists for each benchmark problem: the name of the problem (Problem);
the type of conjecture (Q), i.e., whether the conjecture is existential $\exists$ or universal $\forall$; the status of the conjecture (Status), i.e.,
true if the conjecture is a consequence and false otherwise; the maximum number of variables in any clause ($X$); the number of variables in the conjecture ($Y$);
the number of test points/constants introduced by the Hammer ($B$); the size of the formula in kilobyte before and after the hammering (Size);
the total time (in s) needed by our toolchain to solve the problem (t-time); the time (in s) spent on hammering the input formula (h-time);
the time (in s) spent on parsing the hammered formula by \Rulewerk{} (p-time); the time (in s) \Rulewerk{} actually spent on reasoning (r-time). 
The remaining four columns list the time in s needed by the other tools to solve the benchmark problems. 
An entry "N/A" means that the benchmark example cannot be expressed in the tools input format, e.g., it is not possible to encode a universal conjecture (or, to be more precise, its negation) in the CHC format. 
An entry "timeout" means that the tool could not solve the problem in the given time limit of 40 minutes.
Rulewerk is connected to SPASS-SPL via a file interface. Therefore, we show parsing time separately.

The experiments show that only our toolchain solves all the problems in reasonable time.
It is also the only solver that can decide in reasonable time whether a universal conjecture is not a consequence.
This is not surprising because to our knowledge our toolchain is the only theorem prover that implements a decision procedure for \HBS(\SB).
On the other types of problems, our toolchain solves all of the problems in the range of seconds and with comparable times to the best tool for the problem.
For problems with existential conjectures, the CHC solver spacer is the best, but as a trade-off it is unable to handle universal conjectures.
The instantiation techniques employed by cvc4 are good for proving some universal conjectures, 
but both SMT solvers seem to be unable to disprove conjectures. Vampire performed best on the hammered problems among all first-order theorem provers we tested,
including iProver~\cite{Korovin08}, E~\cite{SchulzEtAl19}, and SPASS~\cite{WeidenbachEtAlSpass2009}. We tested all provers in default theorem
proving mode, but adjusted the memory limit of Vampire, because it ran out of memory on ecu\_u4 with the default setting. The experiments with
the first-order provers showed that our hammer also works reasonably well for them, e.g., they can all solve all lane change problems in less
than a second, but they are simply not specialized for the $\HBS$ fragment.


\section{Conclusion} \label{sec:conclusion}

We have presented several new techniques that allow us to translate $\BS(\SB)\PP$
clause sets with both universally and existentially quantified conjectures into logics for which efficient decision procedures exist.
The first set of translations returns a finite abstraction for our clause set and conjecture, i.e.,
an equisatisfiable ground $\BS(\LRA)$ clause set over a finite set of test points/constants that can be solved in theory by any SMT solver for linear arithmetic.
The abstraction grows exponentially in the maximum number of variables in any input clause.
Realistic supervisor examples have clauses with 10 or more variables and the basis of the growth exponent is also typically large, e.g., in our examples it ranges from 15 to 1500,
so this leads immediately to very large clause sets. An exponential growth in grounding is also unavoidable, because the abstraction reduces
a NEXPTIME-hard problem to an NP-complete problem (ground $\BS$, i.e., SAT).
As an alternative, we also present a Datalog hammer, i.e., a translation to an equisatisfiable $\HBS$ clause set without any theory constraints. 
The hammer is restricted to the Horn case, i.e., $\HBS(\SB)\PP$ clauses, and the conjectures to positive universal/existential conjectures.
Its advantage is that the formula grows only exponentially in the number of variables in the universal conjecture. 
This is typically much smaller than the maximum number of variables in any input clause, e.g., 
in our examples it never exceeds three.

We have implemented the Datalog hammer into our $\BS(\LRA)$ system SPASS-SPL and combined it with the Datalog reasoner Rulewerk.
The resulting toolchain is an effective way of deciding verification conditions for supervisors if the supervisors can be
modeled as $\HBS(\SB)$ clause sets and the conditions as positive $\BS(\SB)$ conjectures.
To confirm this, we have presented two use cases for real-world supervisors: (i)~the verification of supervisor code for the electrical
control unit of a super-charged combustion engine and (ii)~the continuous certification of lane assistants.
Our experiments show that for these use cases our toolchain is overall superior
to existing solvers. Over existential conjectures, it is comparable with existing solvers (e.g., CHC solvers). Moreover, our toolchain is the only solver we are aware of that can proof and disproof universal conjectures for our use cases.

For future work, we want to further develop  our toolchain in several directions. 
First, we want SPASS-SPL to produce explications that prove that its translations are correct.
Second, we plan to exploit specialized Datalog expressions and techniques
(e.g., aggregation and stratified negation) to increase the efficiency of our toolchain and to lift some restrictions from our input formulas.
Third, we want to optimize the selection of test points. 
For instance, we could partition all predicate argument positions into independent sets, i.e.,
two argument positions are dependent if they are assigned the same variable in the same rule. For each of these partitions,
we should be able to create an independent and much smaller set of test points because we only have to consider theory constraints connected
to the argument positions in the respective partition. In many cases, this would lead to much smaller sets of test points
and therefore also to much smaller hammered and finitely abstracted formulas.


\smallskip\noindent
{\bf Acknowledgments:} This work was funded by DFG grant 389792660 as part of
\href{http://perspicuous-computing.science}{TRR~248 (CPEC)},
by BMBF in project \href{https://www.scads.de}{ScaDS.AI},
and by the \href{https://cfaed.tu-dresden.de/}{Center for Advancing Electronics Dresden} (cfaed).
We thank Pascal Fontaine, Alberto Griggio, Andrew Reynolds, Stephan Schulz and our anonymous reviewers
for discussing various aspects of this paper.

\if 1\IsExtended
\bibliographystyle{acm}
\else
\bibliographystyle{splncs04}
\fi
\bibliography{paper}

\end{document}